\documentclass[a4paper,fleqn,usenatbib]{mnras}
\usepackage{txfonts}
\usepackage{graphicx}

\title[A millisecond pulsar in an extremely wide binary system]{A millisecond
  pulsar in an extremely wide binary system}

\author[Bassa et al.]  {C.\,G.\,Bassa$^1$\thanks{email: bassa@astron.nl},
  G.\,H.\,Janssen$^1$,
  B.\,W.\,Stappers$^2$,
  T.\,M.\,Tauris$^{3,4}$,
  T. Wevers$^5$,
  P.\,G.\,Jonker$^{6,5}$,\newauthor
  L.\,Lentati$^7$,
  J.\,P.\,W.\,Verbiest$^{8,3}$,
  G.\,Desvignes$^3$,
  E.\,Graikou$^3$,
  L.\,Guillemot$^{9,10}$,
  P.\,C.\,C.\,Freire$^3$,\newauthor
  P.\,Lazarus$^3$,
  R.\,N.\,Caballero$^3$,
  D.\,J.\,Champion$^3$,
  I.\,Cognard$^{9,10}$,
  A.\,Jessner$^3$,
  C.\,Jordan$^2$,\newauthor
  R.\,Karuppusamy$^3$,
  M.\,Kramer$^{3,2}$,
  K.\,Lazaridis$^3$,
  K.\,J.\,Lee$^{11,3}$,
  K.\,Liu$^{3,10}$,
  A.\,G.\,Lyne$^2$,\newauthor
  J.\,McKee$^2$,
  S.\,Os{\l}owski$^{8,3}$,
  D.\,Perrodin$^{12}$,
  S.\,Sanidas$^{13}$,
  G.\,Shaifullah$^{8,3}$,
  R.\,Smits$^1$,\newauthor
  G.\,Theureau$^{9,10,14}$,
  C. Tiburzi$^{8,3}$ and
  W.\,W.\,Zhu$^3$\newauthor\\
  $^1$ASTRON, the Netherlands Institute for Radio Astronomy, Postbus 2, 7990 AA, Dwingeloo, the Netherlands\\
  $^2$Jodrell Bank Centre for Astrophysics, The University of Manchester, Manchester, M13\,9PL, United Kingdom\\
  $^3$Max-Planck-Institut f\"ur Radioastronomie, Auf dem H\"ugel 69, D-53121 Bonn, Germany\\
  $^4$Argelander-Institut f\"ur Astronomie, Universit\"at Bonn, Auf dem H\"ugel 71, D-53121 Bonn, Germany\\
  $^5$Department of Astrophysics/IMAPP, Radboud University Nijmegen, P.O. box 9010, 6500 GL Nijmegen, The Netherlands\\
  $^6$SRON, Netherlands Institute for Space Research, Sorbonnelaan 2, 3584 CA Utrecht, the Netherlands\\
  $^7$Institute of Astronomy / Battcock Centre for Astrophysics, University of Cambridge, Madingley Road, Cambridge CB3 0HA, United Kingdom\\
  $^8$Fakult\"at f\"ur Physik, Universit\"at Bielefeld, Postfach 100131, 33501 Bielefeld, Germany\\
  $^9$Laboratoire de Physique et Chimie de l'Environnement et de l'Espace LPC2E CNRS-Universit{\'e} d'Orl{\'e}ans, F-45071 Orl{\'e}ans, France\\
  $^{10}$Station de radioastronomie de Nan{\c c}ay, Observatoire de Paris, CNRS/INSU F-18330 Nan{\c c}ay, France\\
  $^{11}$Kavli institute for astronomy and astrophysics, Peking University, Beijing 100871, P.R. China\\
  $^{12}$INAF - Osservatorio Astronomico di Cagliari, via della Scienza 5, I-09047 Selargius (CA), Italy\\
  $^{13}$Anton Pannekoek Institute for Astronomy, University of Amsterdam, Science Park 904, 1098 XH Amsterdam, The Netherlands\\
  $^{14}$Laboratoire Univers et Th\'eories, Observatoire de Paris, CNRS/INSU, Universit\'e Paris Diderot, 5 place Jules Janssen, 92190 Meudon, France\\
}

\date{Accepted \today. Received \today; in original form \today}

\pubyear{2016}

\begin{document}
\label{firstpage}
\pagerange{\pageref{firstpage}--\pageref{lastpage}} 
\maketitle

\begin{abstract}
  We report on 22\,yrs of radio timing observations of the millisecond
  pulsar J1024$-$0719 by the telescopes participating in the European
  Pulsar Timing Array (EPTA). These observations reveal a significant
  second derivative of the pulsar spin frequency and confirm the
  discrepancy between the parallax and Shklovskii distances that has
  been reported earlier. We also present optical astrometry,
  photometry and spectroscopy of 2MASS\,J10243869$-$0719190. We find
  that it is a low-metallicity main-sequence star (K7V spectral type,
  $\mathrm{[M/H]}=-1.0$, $T_\mathrm{eff}=4050\pm50$\,K) and that its
  position, proper motion and distance are consistent with those of
  PSR\,J1024$-$0719. We conclude that PSR\,J1024$-$0719 and
  2MASS\,J10243869$-$0719190 form a common proper motion pair and are
  gravitationally bound. The gravitational interaction between the
  main-sequence star and the pulsar accounts for the spin frequency
  derivatives, which in turn resolves the distance discrepancy. Our
  observations suggest that the pulsar and main-sequence star are in
  an extremely wide ($P_\mathrm{b}>200$\,yr) orbit. Combining the
  radial velocity of the companion and proper motion of the pulsar, we
  find that the binary system has a high spatial velocity of
  $384\pm45$\,km\,s$^{-1}$ with respect to the local standard of rest
  and has a Galactic orbit consistent with halo objects. Since the
  observed main-sequence companion star cannot have recycled the
  pulsar to millisecond spin periods, an exotic formation scenario is
  required. We demonstrate that this extremely wide-orbit binary could
  have evolved from a triple system that underwent an asymmetric
  supernova explosion, though find that significant fine-tuning during
  the explosion is required. Finally, we discuss the implications of
  the long period orbit on the timing stability of PSR\,J1024$-$0719
  in light of its inclusion in pulsar timing arrays.
\end{abstract}

\begin{keywords}
  stars: individual: PSR\,J1024$-$0719 -- stars: neutron -- binaries:
  general -- superovae: general
\end{keywords}

\section{Introduction}
The recent direct detection of gravitational waves by ground-based
interferometers has opened a new window on the Universe
\citep{aaa+16}. Besides ground-based interferometers, gravitational
waves are also predicted to be measurable using a pulsar timing array
(PTA), in which an ensemble of radio millisecond pulsars are used as
extremely stable, celestial clocks representing the arms of a Galactic
gravitational wave detector \citep{det79,hd83}.

At present, three PTAs are in operation; the European Pulsar Timing
Array in Europe (EPTA; \citealt{dcl+16}), the Parkes Pulsar Timing
Array (PPTA) in Australia \citep{mhb+13} and NANOGrav in North-America
\citep{dfg+13}. The three PTAs have been in operation for about a
decade, and have so far provided limits on the stochastic
gravitational wave background at nano-hertz frequencies
(e.g.\ \citealt{ltm+15,src+13,abb+15}). The three PTAs collaborate
together as the International Pulsar Timing Array (IPTA;
\citealt{vlh+16}). As a result of improvements in instrumentation,
analysis software, and higher observing cadence, the timing precision
has increased with time, which leads previously unmodelled effects to
become important.

The first data release by the EPTA (EPTA DR1.0) has recently been
published by \citet{dcl+16} and provides high-precision pulse
times-of-arrival (TOAs) and timing ephemerides for an ensemble of 42
radio millisecond pulsars spanning baselines of 7 to 18\,yrs in
time. While preparing this data release, we rediscovered the apparent
discrepancy in the distance of the isolated millisecond pulsar
J1024$-$0719 which was reported by \citet{egc+13} and also noticed by
\citet{mnf+16}. This discrepancy arises due to the high proper motion
of the pulsar, which gives rise to an apparent positive radial
acceleration and hence a change in the spin period known as the
Shklovskii effect \citep{shk70}. In the case of PSR\,J1024$-$0719, the
Shklovskii effect exceeds the observed spin period derivative for
distances larger than 0.43\,kpc. Pulsar timing yields parallax
distances beyond that limit, which would require an unphysical
negative intrinsic spin period derivative for PSR\,J1024$-$0719.

This discrepancy could be resolved through a negative radial
acceleration due to a previously unknown binary companion in a very
wide orbit. In a search for non-thermal emission from isolated
millisecond pulsars, \citet{srr+03} studied the field of
PSR\,J1024$-$0719 in optical bands. Their observations revealed the
presence of two stars near the position of the pulsar. The broadband
colors and spectrum of the bright star, 2MASS\,J10243869$-$0719190
($R\sim19$, hereafter star B) were consistent with those of a
main-sequence star of spectral type K5, while the other, fainter star
($R\sim24$, hereafter star F) had broadband colors which
\citet{srr+03} suggested could be consistent with non-thermal emission
of the neutron star in PSR\,J1024$-$0719. However, the astrometric
uncertainties in both the optical and radio did not allow
\citet{srr+03} to make a firm conclusion on the association of either
stars B or F and PSR\,J1024$-$0719.

Here, we present radio observations of PSR\,J1024$-$0719 and optical
observations of stars B and F (\S\ref{sec:observations}) that allow us
to improve the timing ephemeris of PSR\,J1024$-$0719 and the
astrometry, photometry and spectroscopy of stars B and/or F
(\S\ref{sec:analysis}). In \S\ref{sec:results} we present our results
which show that star B is in an extremely wide orbit around
PSR\,J1024$-$0719. An evolutionary formation scenario for this wide
orbit is presented in \S\,\ref{sec:formation}. We discuss our findings
and conclude in \S\ref{sec:conclusions}. This research is the result
of the common effort to directly detect gravitational waves using
pulsar timing, known as the European Pulsar Timing Array
\citep{dcl+16}\footnote{\url{http://www.epta.eu.org}}.

While preparing this paper, we became aware that a different group had
used independent radio timing observations and largely independent
optical observations to reach the same conclusion about the binarity
of PSR\,J1024$-$0719 \citep{kkn+16}. The analysis and results
presented in this paper agree very well with those of \citet{kkn+16},
though these authors suggest an alternate formation scenario for
PSR\,J1024$-$0719 (see \S\ref{ssec:alternative}).

\section{Observations}\label{sec:observations}
\subsection{Radio}
We present radio timing observations of PSR\,J1024$-$0719 that were
obtained with telescopes participating in the EPTA over the past
22\,years. The EPTA DR1.0 data presented by \citet{dcl+16} contains
timing observations of PSR\,J1024$-$0719 from the 'historical' pulsar
instrumentation at Effelsberg, Jodrell Bank, Nan\c cay and Westerbork
and contains 561 times-of-arrival (TOAs) over a 17.3\,yr time span
from January 1997 to April 2014.

The DR1.0 data set was extended back to March 1994 with observations
obtained at 410, 610 and 1400\,MHz with the Jodrell Bank analog
filterbank (AFB; \citealt{sl96}). The first year of this data set has
been published in the PSR\,J1024$-$0719 discovery paper by
\citet{bjb+97}. We also included observations obtained with the new
generation of baseband recording/coherent dedispersing pulsar-timing
instruments presently in operation within the EPTA. These instruments
are \textsc{PuMa\,II} at Westerbork \citep{kss08}, \textsc{PSRIX} at
Effelsberg \citep{lkg+16}, the \textsc{ROACH} at Jodrell Bank
\citep{bjk+16} and \textsc{BON} and \textsc{NUPPI} at Nan\c cay
\citep{ct06,ctg+13}. A subset of the Nan\c cay TOAs on
PSR\,J1024$-$0719 were presented in \citet{gsl+16}. Combining these
data sets yields 2249\,TOAs spanning 22\,yrs since the discovery of
PSR\,J1024$-$0719 (Table\,\ref{tab:ephemeris} and
Fig.\,\ref{fig:residuals}).

\subsection{Optical}
We retrieved imaging observations of the field of PSR\,J1024$-$0719
from the ESO archive. These observations consist of $3\times2$\,min
$V$-band, $3\times3$\,min $R$-band, and $3\times2$\,min $I$-band
exposures, which were obtained with the FORS1 instrument
\citep{aff+98} at the ESO VLT at Cerro Paranal on March 28, 2001 under
clear conditions with $0\farcs6$ seeing. The high resolution
collimator was used, providing a $0\farcs1$\,pix$^{-1}$ pixel scale
and a $3\farcm4\times3\farcm4$ field-of-view. Photometric standards
from the \citet{lan92} SA109 field were observed with the same filters
and instrument setup. Analysis of this data has been presented by
\citet{srr+03}.

The field of PSR\,J1024$-$0719 was also observed with OmegaCAM
\citep{kui11} at the VLT Survey Telescope. OmegaCAM consists of 32
4k$\times$2k CCDs with a $0\farcs21$\,pix$^{-1}$ pixel scale. SDSS
$ugriz$ observations were obtained between February 24 and March 2,
2012 as part of the ATLAS survey \citep{smc+15}. Here we specifically
use the 45\,s SDSS $r$-band exposure obtained on February 27,
2012. The seeing of that exposure was $0\farcs75$.

Follow-up observations of the field of PSR\,J1024$-$0719 were taken on
June 10 and 11, 2015 with FORS2 at the ESO VLT at Cerro Paranal in
Chile. A series of 8 dithered 5\,s exposures in the $R$-band filtered
were obtained on June 10th, under clear conditions with mediocre
seeing of $1\farcs1$. The standard resolution collimator was used with
$2\times2$\,binning, providing a pixel scale of $0\farcs25$\,pix$^{-1}$
with a $6\farcm8\times6\farcm8$ field-of-view.

Two long-slit spectra were also obtained with FORS2, one with the
600RI grism on June 10th and one with the 600z grism on June
11th. These grisms cover wavelength ranges of 5580 to 8310\,\AA\ and
7750 to 10430\,\AA, respectively. Both exposures were 1800\,s in
length and used the $1\farcs0$ slit. The CCDs were read out using
$2\times2$ binning, providing a resolution of 6.3\,\AA, sampled at
1.60\,\AA\,pix$^{-1}$ for both the 600RI and 600z grism. The seeing
during these exposures was $0\farcs77$ and $0\farcs80$,
respectively. The spectrophotometric standard LTT\,3864 was observed
with the same grisms and a $5\arcsec$ slit. Arc-lamp exposures for all
grism and slit combinations were obtained as part of the normal VLT
calibration programme.

\begin{table}
  \caption{Parameters for PSR J1024$-$0719. All reported uncertainties
    have been multiplied by the square root of the reduced $\chi^2$.}
  \label{tab:ephemeris}
  \begin{tabular}{lc}
    \hline\hline
    \multicolumn{2}{c}{Fit and data-set} \\
    \hline
    MJD range & 49416.8---57437.0 \\
    Data span (yr) & 21.96 \\
    Number of TOAs & 2249 \\
    Rms timing residual ($\mu$s) & 2.81 \\
    Weighted fit &  Y \\
    Reduced $\chi^2$ value  & 0.98 \\
    Reference epoch (MJD) & 55000 \\
    \hline
    \multicolumn{2}{c}{Measured Quantities} \\
    \hline
    Right ascension, $\alpha_\mathrm{J2000}$ & $10^\mathrm{h}24^\mathrm{m}38\fs675380(5)$ \\
    Declination, $\delta_\mathrm{J2000}$ & $-07\degr19\arcmin19\farcs43396(14)$ \\
    Proper motion in R.A., $\mu_{\alpha} \cos \delta$ (mas\,yr$^{-1}$) & $-35.255(19)$ \\
    Proper motion in decl., $\mu_{\delta}$ (mas\,yr$^{-1}$) & $-48.19(4)$ \\
    Parallax, $\pi$ (mas) & 0.77(11) \\
    Pulse frequency, $f$ (s$^{-1}$) & 193.7156834485468(7) \\
    First derivative of $f$, $\dot{f}$ (s$^{-2}$) & $-6.95893(15)\times 10^{-16}$ \\
    Second derivative of $f$, $\ddot{f}$ (s$^{-3}$) & $-3.92(2)\times 10^{-27}$ \\
    Third derivative of $f$, $f^{(3)}$ ($s^{-4}$) & $<2.7\times 10^{-36}$ \\
    Fourth derivative of $f$, $f^{(4)}$ ($s^{-4}$) & $<4.5\times 10^{-44}$ \\
    Dispersion measure, DM (pc\,cm$^{-3}$) & 6.4888(8) \\
    First derivative of DM (pc\,cm$^{-3}$\,yr$^{-1}$) & $3(15)\times 10^{-5}$ \\
    Second derivative of DM (pc\,cm$^{-3}$\,yr$^{-2}$) & $-1.4(1.8)\times 10^{-5}$ \\
    \hline
    \multicolumn{2}{c}{Assumptions} \\
    \hline
    Clock correction procedure & TT(BIPM2011) \\
    Solar system ephemeris model & DE421 \\
    Units & TCB \\
    \hline
  \end{tabular}
\end{table}

\section{Analysis}\label{sec:analysis}

\begin{figure}
  \includegraphics[width=\columnwidth]{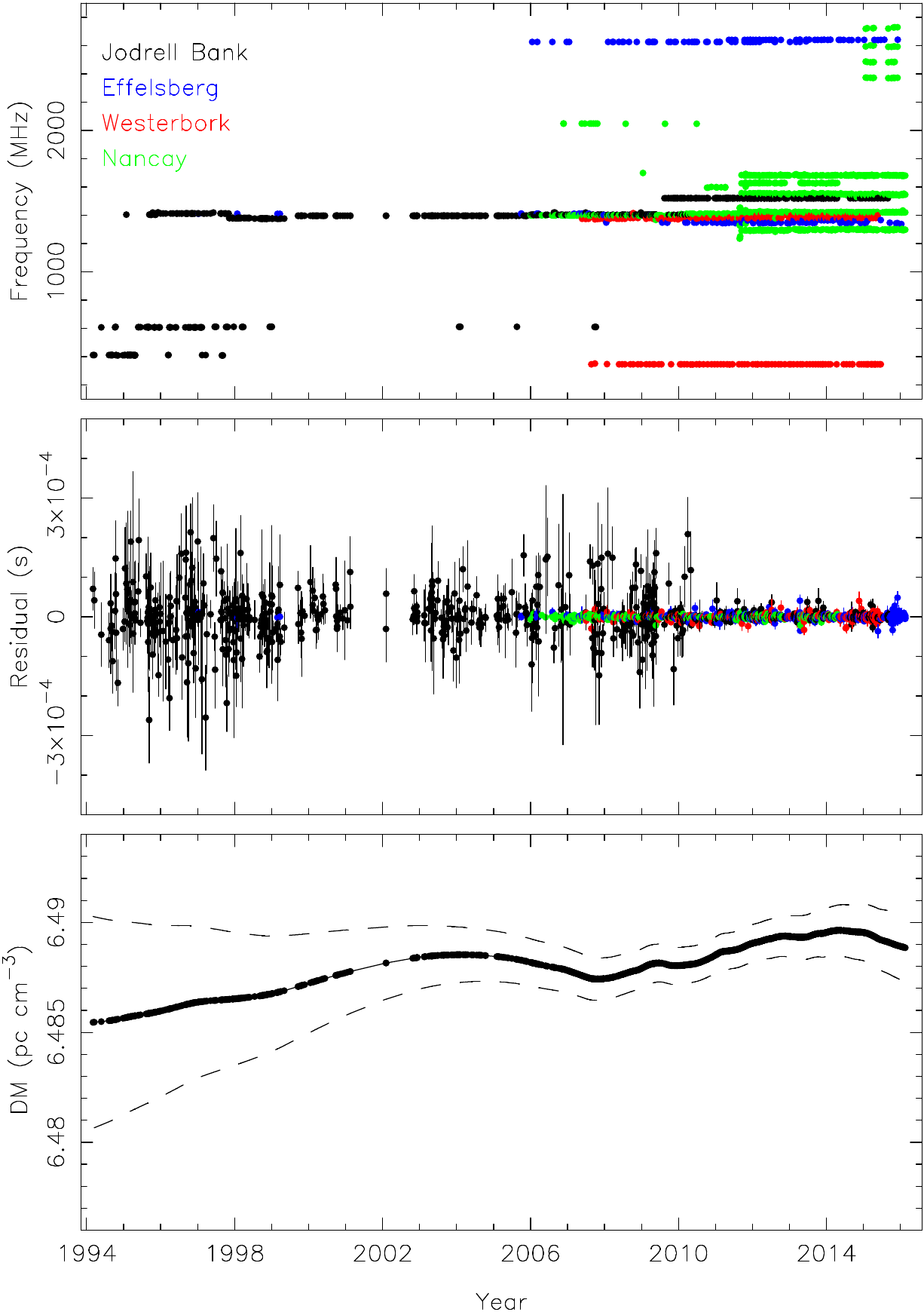}
  \caption{Radio timing observations at different observing
    frequencies are shown in the top panel as a function of
    time. Observations from different observatories are shown in
    different colors. The middle panel shows timing residuals of
    PSR\,J1024$-$0719 based on our timing ephemeris
    (Table\,\ref{tab:ephemeris}). The bottom panel shows the
    dispersion measure (DM) values as modelled by the power law
    model. The dashed lines indicate the uncertainty in the DM. }
  \label{fig:residuals}
\end{figure}

\subsection{Pulsar timing}
We use the standard approach, as detailed in \citet{ehm06}, for
converting the topocentric TOAs to the solar-system barycenter through
the DE421 ephemeris \citep{fwb09} and placing them on the Terrestial
Time standard (BIPM2011;\,\citealt{pet10}). The TOAs from the
different combinations of telescope and instrument were combined in
the \textsc{Tempo2} pulsar timing software package \citep{hem06}.

The best-fit timing solution to the TOAs was found through the
standard \textsc{Tempo2} iterative least-squares minimization. The
free parameters in the fit consist of the astrometric parameters
(celestial position, proper motion and parallax), a polynomial
describing the spin frequency $f$ as a function of time, and a
polynomial to describe variations in the pulsar dispersion measure
(DM) as a function of time. We also fitted for offsets
between the telescope and instrument combinations to take into account
difference in pulse templates and instrument delays.

To perform model selection between models that include increasing
numbers of spin frequency derivatives when incorporating increasingly
complex noise models, we use the Bayesian pulsar timing package
\textsc{TempoNest} \citep{lah+14}, which operates as a plugin to
\textsc{Tempo2}. In all cases we include parameters to model the white
noise that scale, or add in quadrature with, the formal TOA
uncertainties, and define these parameters for each of the 20
observing systems included in the data set.  We also include
combinations of additional time correlated processes, including DM
variations and system-dependent noise.  For this final term we use the
approach described in \cite{lsc+16}.  Apart from the spin frequency,
derivatives we marginalize over the timing model analytically.  The
spin frequency derivatives are included in the analysis numerically
with priors that are uniform in the amplitude of the parameter in all
cases.

When including only white noise parameters, and limiting our model for
DM variations to a quadratic in time, we find the evidence supports a
total of four frequency-derivatives.  However, we find there is
significant support from the data for higher order DM variations.  In
particular, the difference in the log evidence for models with and
without an additional power law DM model besides the quadratic model
is 36 in favor of including the additional stochastic term.  We find
the spectral index of the DM variations ($3.2\pm0.6$) to be consistent
with that expected for a Kolmogorov turbulent medium.  When including
these additional parameters to describe higher-order DM variations, we
find that the data only support the first and second
frequency-derivatives.  The significant detections of the third and
fourth frequency derivatives observed in the simpler model can
therefore not be confidently interpreted as arising from a binary
orbit, as they are highly covariant with the power-law variations in
DM, thereby reducing their significance. Finally, we find no support
for any system dependent time-correlated noise in this data set. The
timing ephemeris using a DM power law model is tabulated in
Table\,\ref{tab:ephemeris} and timing residuals and DM values as a
function of time are plotted in Fig.\,\ref{fig:residuals}.

\subsection{Photometry}
The FORS1 and FORS2 images were bias-corrected and flatfielded using
sky flats. Instrumental magnitudes were determined through point
spread function (PSF) fitting with the \textsc{DAOPHOT II} package
\citep{ste87}. The instrumental PSF magnitudes of the 2001 $V\!RI$
observations were calibrated against 17 photometric standards from the
SA109 field, using calibrated values from \citet{ste00} and fitting
for zeropoint offsets and color terms, but using the tabulated ESO
extinction coefficients of 0.113, 0.109 and 0.087\,mag\,airmass$^{-1}$
for $V\!RI$, respectively. The residuals of the fit were 0.03\,mag in
$V$, 0.06\,mag in $R$ and 0.07\,mag in $I$. We find that star B has
$V=19.81\pm0.04$, $R=18.78\pm0.06$ and $I=18.00\pm0.07$, while star F
has $V=24.71\pm0.13$, $R=24.64\pm0.14$ and $I=24.57\pm0.34$.

The instrumental magnitudes of the 8 FORS2 $R$-band images from 2015
were calibrated against the 2001 FORS1 observations. As the FORS2
$R$-band (\textsc{R\_SPECIAL}) filter has a different transmission
curve compared to the FORS1 Bessell $R$-band (\textsc{R\_BESS})
filter, we fitted $R$-band offsets as a function of the $R-I$ color,
finding a color term of $-0.11(R-I)$ and residuals of 0.05\,mag. We
find that star B has $R=18.74\pm0.05$, i.e.\ consistent with the FORS1
measurement.

\subsection{Astrometry}
We use the OmegaCAM $r$-band exposure to obtain absolute astrometry of
the field of PSR\,J1024$-$0719 as the individual CCDs from OmegaCAM
have the largest field-of-view each ($7\farcm3\times14\farcm5$). The
image containing the field of PSR\,J1024$-$0719 was calibrated against
astrometric standards from the fourth U.S.\ Naval Observatory CCD
Astrograph Catalog (UCAC4, \citealt{zfg+13}), which provides proper
motions through comparison with other catalogs. A total of 14 UCAC4
astrometric standards overlapped with the image. The UCAC4 positions
were corrected for proper motion from the reference epoch (2000.0) to
the epoch of the OmegaCAM image (2012.15). The centroids of these
stars were measured and used to compute an astrometric calibration
fitting for offset, scale and position angle. Rejecting two outliers
with residuals larger than $0\farcs3$, the calibration has rms
residuals of $0\farcs047$ in right ascension and $0\farcs040$ in
declination.

The position of star B on the OmegaCAM $r$-band image is
$\alpha_\mathrm{J2000}=10^\mathrm{h}24^\mathrm{m}38\fs677(3)$,
$\delta_\mathrm{J2000}=-07\degr19\arcmin19\farcs59(4)$, where the
uncertainties are the quadratic sum of the uncertainty in the
astrometric calibration and the positional uncertainty on the image
($0\farcs008$ in both coordinates). Star F is not detected on the
OmegaCAM images. Hence, we transferred the OmegaCAM calibration to the
median combined FORS1 $R$-band image using 7 stars in common to both
(excluding star B). The residuals of this transformation are
$0\farcs06$ in both coordinates. Star F is at
$\alpha_\mathrm{J2000}=10^\mathrm{h}24^\mathrm{m}38\fs691(5)$,
$\delta_\mathrm{J2000}=-07\degr19\arcmin17\farcs45(7)$. Here, the
uncertainties also include the uncertainty in the calibration between
the OmegaCAM and FORS1 image.

The 14.2\,yr time baseline between the FORS1 and FORS2 images allows
us to determine the proper motions of stars B and F. To prevent
pollution by non-random proper motions of stellar objects, we
determined the absolute proper motion with respect to back ground
galaxies. We selected 7 objects on the FORS1 and FORS2 images, located
within $45\arcsec$ of PSR\,J1024$-$0719, which were clearly extended
in comparison to the PSF of stars. The 5\,s FORS2 $R$-band exposures
were not deep enough to accurately measure the positions of the
objects, so instead we used the 40\,s FORS2 acquisition exposure for
the long-slit spectroscopic observations. This image was taken with
the OG590 order sorting filter, which cuts light blue-wards of
6000\,\AA.

Using the 7 extra-galactic objects, we determined the transformation
between each of the three FORS1 $R$-band images and the FORS2 OG590
image. The residuals of the transformations were typically
$0\farcs050$ in each coordinate. The transformations were then used to
compute the positional offsets of stars B and F between the FORS1 and
FORS2 images. Averaging the positional offsets and taking into account
the 14.2\,yr time baseline yielded the proper motion of both stars. We
find that star B had $\mu_\alpha \cos
\delta=-0\farcs033(2)$\,yr$^{-1}$ and
$\mu_\delta=-0\farcs050(2)$\,yr$^{-1}$, while star F has $\mu_\alpha
\cos \delta=0\farcs001(5)$\,yr$^{-1}$ and
$\mu_\delta=0\farcs021(5)$\,yr$^{-1}$. Table\,\ref{tab:properties}
lists the magnitudes, position and proper motion of stars B and F.

\begin{table*}
  \centering
  \caption{The positions and proper motions of stars B and F and
    PSR\,J1024$-$0719 are referenced to epoch MJD\,55984.135. The
    uncertainties in the $V\!RI$ magnitudes of stars B and F are
    instrumental. The zeropoint uncertainties of 0.03\,mag in $V$,
    0.06\,mag in $R$ and 0.07\,mag in $I$ should be added in
    quadrature to obtain absolute uncertainties. }
  \label{tab:properties}
  \begin{tabular}{lllllllll}
    \hline
    Object & $\alpha_\mathrm{J2000}$ & $\delta_\mathrm{J2000}$ & $\mu_\alpha\cos \delta$ (yr$^{-1}$) & $\mu_\delta$ (yr$^{-1}$) & $V$ & $R$ & $I$ & ref.\\
    \hline
    PSR\,J1024$-$0719 & $10^\mathrm{h}24^\mathrm{m}38\fs668996(6)$  & $-07\degr19\arcmin19\farcs56380(18)$ & $-0\farcs035276(18)$ & $-0\farcs04822(4)$ & $\ldots$ & $\ldots$ & $\ldots$ & 1 \\
    PSR\,J1024$-$0719 & $10^\mathrm{h}24^\mathrm{m}38\fs668988(7)$  & $-07\degr19\arcmin19\farcs5638(2)$ & $-0\farcs03528(3)$ & $-0\farcs04818(7)$ & $\ldots$ & $\ldots$ & $\ldots$ & 2 \\
    PSR\,J1024$-$0719 & $10^\mathrm{h}24^\mathrm{m}38\fs668985(13)$ & $-07\degr19\arcmin19\farcs5641(4)$ & $-0\farcs03533(4)$ & $-0\farcs04832(8)$ & $\ldots$ & $\ldots$ & $\ldots$ & 3 \\
    PSR\,J1024$-$0719 & $10^\mathrm{h}24^\mathrm{m}38\fs668996(10)$ & $-07\degr19\arcmin19\farcs5638(3)$ & $-0\farcs0352(1)$ & $-0\farcs0480(2)$ & $\ldots$ & $\ldots$ & $\ldots$ & 4 \\[0.5em]
    Star B (PPMXL) & $10^\mathrm{h}24^\mathrm{m}38\fs674(9)$ & $-07\degr19\arcmin19\farcs54(13)$ & $-0\farcs030(6)$ & $-0\farcs048(6)$ & $\ldots$ & $\ldots$ & $\ldots$ & 5\\
    Star B (APOP)  & $10^\mathrm{h}24^\mathrm{m}38\fs663(3)$ & $-07\degr19\arcmin19\farcs51(3)$  & $-0\farcs0335(19)$ & $-0\farcs0517(12)$ & $\ldots$ & $\ldots$ & $\ldots$ & 6 \\
    Star B & $10^\mathrm{h}24^\mathrm{m}38\fs677(3)$ & $-07\degr19\arcmin19\farcs59(4)$ & $-0\farcs033(2)$ & $-0\farcs050(2)$ & $19.81(2)$ & $18.78(1)$ & $18.00(1)$ & 1 \\[0.5em]
    Star F & $10^\mathrm{h}24^\mathrm{m}38\fs692(6)$ & $-07\degr19\arcmin17\farcs22(9)$ & $-0\farcs002(5)$ & $+0\farcs008(5)$ & $24.71(13)$ & $24.64(13)$ & $24.6(3)$ & 1 \\
    \hline
  \end{tabular}
  References: (1) this work; (2) \cite{dcl+16}; (3) \cite{rhc+16}; (4) \cite{mnf+16};\\ (5) \cite{rds10}; (6) \citet{qyb+15}
\end{table*}

\subsection{Spectroscopy}
The spectroscopic observations were bias corrected and flatfielded
using lamp-flats. Spectra of the counterpart and the
spectrophotometric standard were extracted using the optimal
extraction method by \citet{hor86}, and wavelength calibrated using
the arc-lamp exposures. The residuals of the wavelength calibration
were $0.1$\,\AA\ or better. The extracted instrumental fluxes were
corrected for slit losses by using the wavelength dependent spatial
profile to estimate the fraction of the flux being masked by the
finite slit width (about 28\% for the 1$\arcsec$ slit and 1\% for the
5$\arcsec$ slit). The instrumental response of both instrument setups
was determined from the standard observations and a comparison with
calibrated spectra from \citet{hws+92,hsh+94}. The response was then
used to flux calibrate the spectra of star B.

\section{Results}\label{sec:results}
\subsection{Pulsar timing}
As the data set presented here is an extension of the EPTA DR1.0
data set, the parameters in Table\,\ref{tab:ephemeris} have improved
with respect to the \citet{dcl+16} ephemeris. The parameters are
generally consistent with the exception of the polynomial describing
the DM variations, primarily due to the inclusion of the power law
model. The inclusion of the multi-frequency observations from the
Jodrell Bank AFB and PuMa\,II at WSRT have significantly improved the
measurement of DM variations.

Our timing ephemeris is also generally consistent with the PPTA and
NANOGrav results. In Table\,\ref{tab:properties} we list the positions
and proper motion of PSR\,J1024$-$0719 from the timing ephemerides by
\citet{dcl+16,rhc+16} and \citet{mnf+16} propagated to the epoch of
the OmegaCAM observations. We find that the positions and proper
motions are consistent within the uncertainties. The observed parallax
$\pi=0.77(11)$\,mas corresponds to a distance of
$d=1.20^{+0.18}_{-0.14}$\,kpc after correcting for Lutz-Kelker bias
\citep{lk73,vlm10}, and is consistent with the values found by
\citet{dcl+16,rhc+16,mnf+16}.

The improvements in the timing ephemeris allow us to measure a
significant second derivative of the spin frequency $f$;
$\ddot{f}$. This higher order derivative likely accounts for the steep
red noise spectrum seen in the PPTA observations of PSR\,J1024$-$0719
\citep{rhc+16}. Our measurement of $\ddot{f}$ translates to
$\ddot{P}=1.05(6)\times10^{-31}$\,s$^{-1}$. For comparison,
\citet{mnf+16} use the NANOGrav data and the $\dot{P}$ measurement by
\citet{vbc+09} to set a limit on
$\ddot{P}<3.2\times10^{-31}$\,s$^{-1}$. \citet{gsl+16} measured
$\ddot{P}=0.70(6)\times10^{-31}$\,s$^{-1}$ from Nan\c cay observations
of PSR\,J1024$-$0719. This value is inconsistent with our value; we
attribute this difference to unmodelled DM variations in the
\citet{gsl+16} timing solution.

The observed spin frequency derivative $\dot{f}_\mathrm{obs}$ contains
contributions from the intrinsic pulsar spindown, the Shklovskii
effect due to non-zero proper motion \citep{shk70}, differential
Galactic rotation and Galactic acceleration (e.g.\ \citealt{nt95}) and
any contribution due to radial acceleration. These effects are more
commonly described as derivatives of the spin period $P$ such that
$\dot{P}_\mathrm{obs}=\dot{P}_\mathrm{int}+\dot{P}_\mathrm{shk}+\dot{P}_\mathrm{dgr}+\dot{P}_\mathrm{kz}+\dot{P}_\mathrm{acc}$.

\begin{figure}
  \includegraphics[width=\columnwidth]{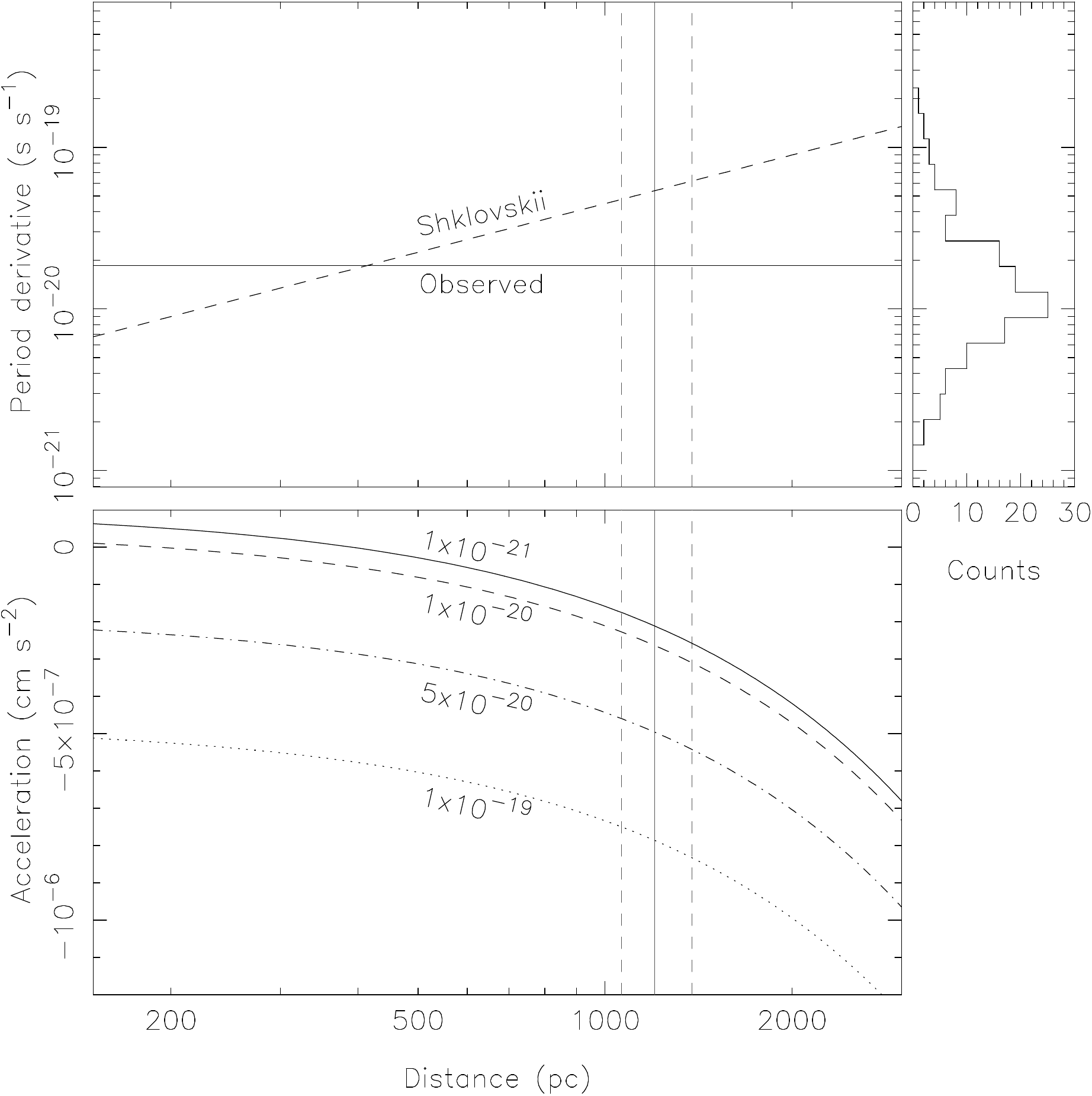}
  \caption{The top panel shows the spin period derivative due to the
    non-zero proper motion (the Shklovskii effect) as a function of
    distance as the dashed line. For distances larger than 0.43\,kpc,
    it exceeds the observed spin period derivative (solid horizontal
    line). The histogram on the right-hand side of the top panel shows
    the distribution of intrinsic spin period derivatives for all
    radio millisecond pulsars in the Galactic field
    \citep{mhth05}. The bottom panel shows the radial acceleration
    required to explain the observed spin period derivative of
    PSR\,J1024$-$0719 given the Shklovskii contribution assuming four
    different values for the intrinsic spin period derivative of the
    pulsar. The solid and dashed vertical lines denote the
    \citet{lk73} bias corrected distance of PSR\,J1024$-$0719 derived
    from our parallax measurement.}
  \label{fig:distance}
\end{figure}

Figure\,\ref{fig:distance} shows the observed spin period derivative
and the Shklovskii contribution as a function of distance. As noted by
\citet{dcl+16}, \citet{mnf+16} and \citet{gsl+16}, the large proper
motion of PSR\,J1024$-$0719 leads the Shklovskii contribution
$\dot{P}_\mathrm{shk}/P=\mu^2 d/c$ to exceed the observed spin period
derivative $\dot{P}_\mathrm{obs}=1.855\times10^{-20}$\,s\,s$^{-1}$ for
distances larger than 0.43\,kpc. Our results confirm this. The
histogram on the right of Fig.\,\ref{fig:distance} shows the
distribution of intrinsic spin period derivatives
$\dot{P}_\mathrm{int}$ of known millisecond pulsars ($P<10$\,ms;
\citealt{mhth05}). These range between $10^{-21}$ and
$10^{-19}$\,s\,s$^{-1}$, and PSR\,J1024$-$0719 is expected to have a
$\dot{P}_\mathrm{int}$ in this range. \citet{dcl+16} and
\citet{mnf+16} estimate that for PSR\,J1024$-$0719 the contributions
due to differential Galactic rotation and Galactic acceleration only
account for
$\dot{P}_\mathrm{dgr}+\dot{P}_\mathrm{kz}<-8\times10^{-22}$\,s\,s$^{-1}$,
which is negligible with respect to the other terms making up
$\dot{P}_\mathrm{obs}$. Depending on the value of
$\dot{P}_\mathrm{int}$, the remaining spin period derivative due to
radial acceleration $\dot{P}_\mathrm{acc}=\ddot{z}_1P_0/c$ required to
explain the observed spin period derivative sets the radial
acceleration $\ddot{z}_1$ between $-2\times10^{-7}$ and
$-8\times10^{-7}$\,cm\,s$^{-2}$.

\subsection{Association with PSR\,J1024$-$0719}
In Table\,\ref{tab:properties}, we list the proper motion and pulsar
position propagated to the epoch of the OmegaCAM observations (MJD
55984.135) for several timing ephemerides of PSR\,J1024$-$0719. Within
the uncertainties, the pulsar position and proper motion as measured
by the three pulsar timing arrays are consistent with each other. We
find that star B is offset from PSR\,J1024$-$0719 by $\Delta
\alpha=-0\farcs12(5)$ and $\Delta\delta=-0\farcs03(4)$, corresponding
to a total offset of $0\farcs12(6)$. Here, the uncertainty is
dominated by the astrometric calibration against the UCAC4
catalog. For star F, the total offset is $2\farcs37(11)$. Moreover,
the proper motions determined from the 14.2\,year baseline between the
FORS1 and FORS2 observations show that, within $2\sigma$, star B has a
proper motion consistent with the pulsar.

\begin{figure}
  \includegraphics[width=\columnwidth]{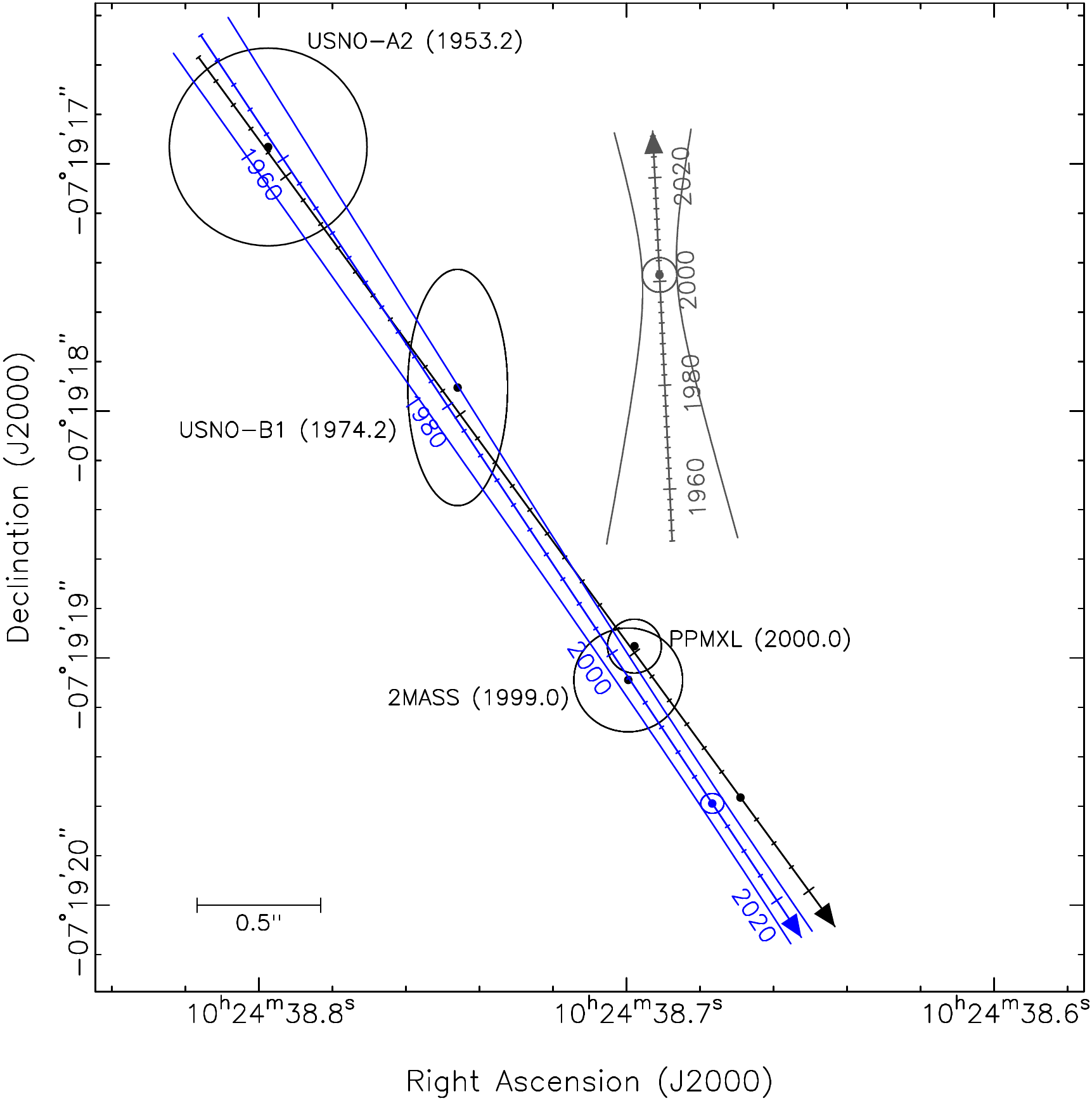}
  \caption{The position of PSR\,J1024$-$0719 (black), star B (blue)
    and star F (grey) between 1950 and 2025. The positional
    uncertainties of stars B and F are indicated by the ellipses at
    the OmegaCAM observation of star B (epoch 2012.15) and the FORS1
    observation of star F (epoch 2001.24). The lines expanding from
    the ellipse illustrate the positional uncertainty due to position
    and proper motion as a function of time. On the scale of this
    plot, the uncertainties on the timing position and proper motion
    of PSR\,J1024$-$0719 are negligible. Also plotted are measurements
    of the position of star B in the USNO-A2 \citep{mbc+98}, USNO-B1
    \citep{mlc+03}, 2MASS \citep{csd+03,scs+06} and PPMXL
    \citep{rds10} catalogs. The epochs of these measurements are given
    in brackets. The USNO-A2 catalog does not provide positional
    uncertainties; we conservatively estimate $0\farcs4$ uncertainties
    on the position.  }
  \label{fig:pos}
\end{figure}

\begin{figure}
  \includegraphics[width=\columnwidth]{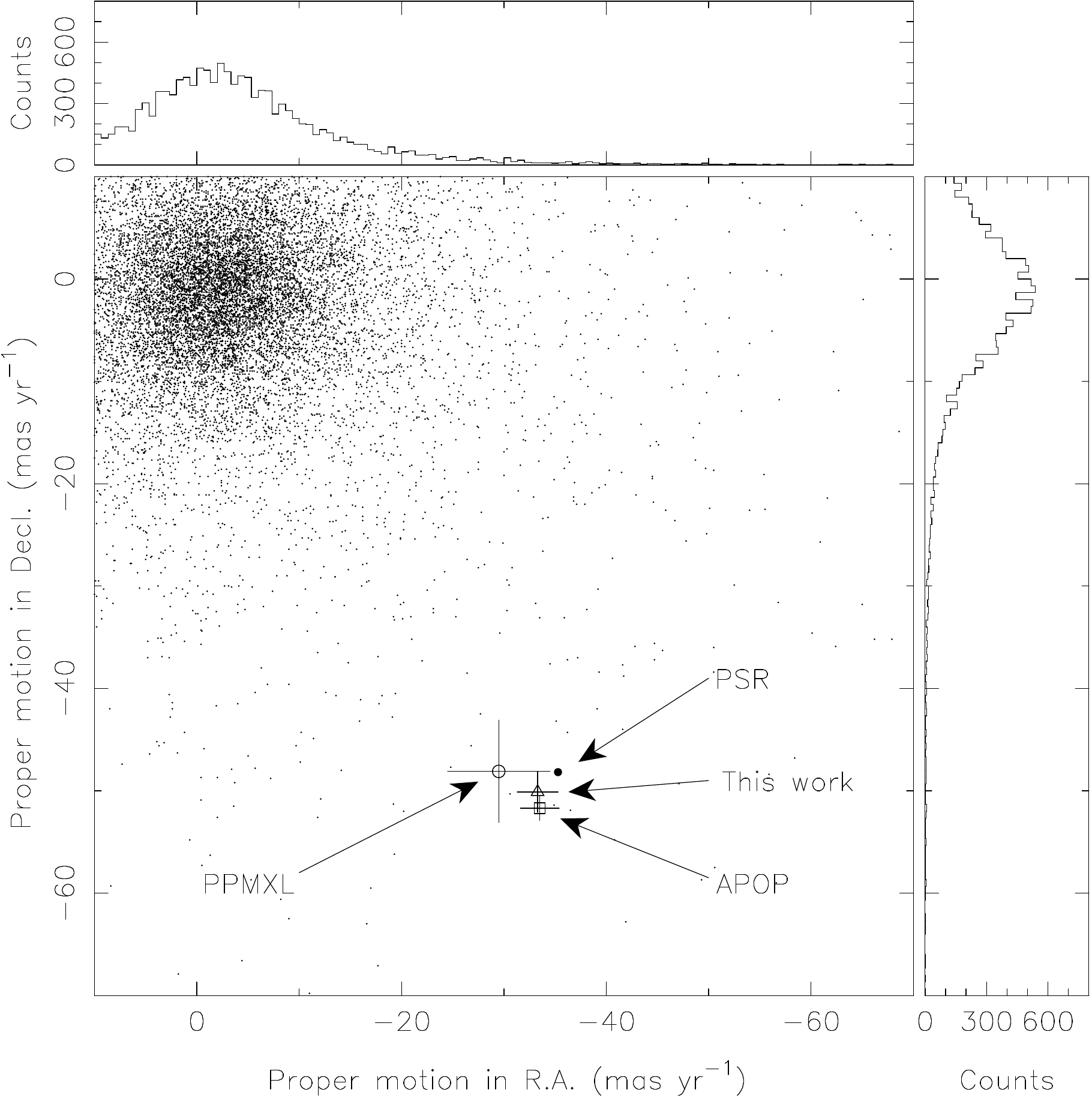}
  \caption{Proper motion measurements of PSR\,J1024$-$0719 and star
    B. The proper motion of PSR\,J1024$-$0719 is denoted by the thick
    black dot. The proper motion of star B as determined in this work,
    the PPMXL catalog \citep{rds10} and the APOP catalog
    \citep{qyb+15} are shown with the triangle, circle and square,
    respectively. The small points and histograms at the top and right
    of the figure represent proper motion measurements from stars in
    the APOP catalog, selecting 14754 stars within a radius of
    $1\degr$ around PSR\,J1024$-$0719. }
  \label{fig:pm}
\end{figure}

The similarity in position and proper motion between star B and
PSR\,J1024$-$0719 is independently confirmed by several surveys. At
$R=18.78$, star B is bright enough to have been recorded on historic
photographic plates and hence is included in the USNO-A2
\citep{mbc+98} and USNO-B1 \citep{mlc+03} astrometric catalogs. It is
also detected in the near-IR in the 2MASS survey
\citep{csd+03,scs+06}. Star B is present in the PPMXL catalog by
\citet{rds10}, which combines the USNO-B1 and 2MASS astrometry to
determine proper motions, and also the APOP catalog by \citet{qyb+15},
which uses STScI digitized Schmidt survey plates originally utilized
for the creation of the GSC\,II catalog \citep{llm+08} to derive
absolute proper motions. The position and proper motion of star B in
these catalogs is plotted in Fig.\,\ref{fig:pos} and
Fig.\,\ref{fig:pm} and listed in Table\,\ref{tab:properties}.

The probability that an unrelated star has, within the uncertainties,
a position and proper motion consistent with PSR\,J1024$-$0719 is
minuscule. The APOP catalog and the FORS2 photometry yield a stellar
density for stars with $R<18.78$ (equal or brighter than star B) of 1
to 2 stars per square arcminute, while within $1\degr$ from
PSR\,J1024$-$0719, only 8 out of 7300 APOP stars with $R<18.78$ have a
proper motion in right ascension and declination that is within
10\,mas\,yr$^{-1}$ of that of PSR\,J1024$-$0719
($\mu=59.71$\,mas\,yr$^{-1}$). Based on these numbers, we estimate that
the chance probability of a star having a similar position and proper
motion to PSR\,J1024$-$0719 is about $10^{-7}$ for stars equal or
brighter than star B. Such a low probability confirms that star B is
associated with PSR\,J1024$-$0719 and that both objects form a common
proper motion pair.

At or above the brightness level of star F, there are about 10 objects
per square arcminute in the FORS1 $R$-band image, suggesting that
there is a probability of about 7\% of finding an object as bright and
close as star F with respect to the pulsar position. Hence, we
consider star F as a field star, not related to PSR\,J1024$-$0719.

\begin{figure*}
  \includegraphics[width=\textwidth]{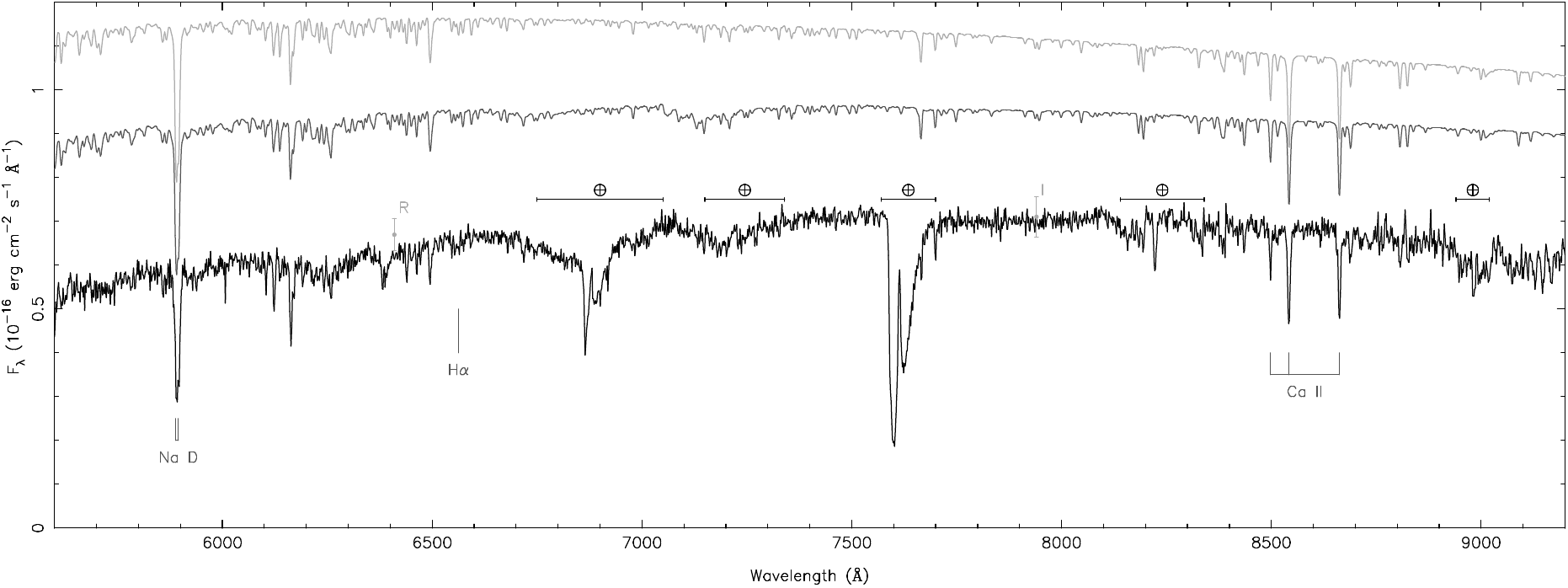}
  \caption{The FORS2 optical spectrum of star B, formed from the
    combination of the single exposures obtained with the 600RI and
    600z grisms, is shown in black. The spectrum has been shifted to
    zero velocity. The $R$ and $I$-band magnitudes have been converted
    to fluxes using the zeropoints of \citet{bcp98} and are
    plotted. Regions affected by telluric absorption have been
    masked. The most prominent spectral lines are indicated. Two
    synthetic spectra by \citet{mscz05} are shown which bracket the
    best fit model to the observed spectrum. Both model spectra have
    $\mathrm{[M/H]=-1.0}$ and has been rotationally broadened by
    83\,km\,s$^{-1}$ and convolved by a truncated Moffat profile
    representing the observational seeing and slit width. The top
    light-grey model spectrum has $T_\mathrm{eff}=4250$\,K and $\log
    g=4.5$\,cgs, while the dark grey model spectrum has
    $T_\mathrm{eff}=4000$\,K and $\log g=4.0$\,cgs. The models have
    been reddened using the extinction law by \citet{ccm89} and
    $E_{B-V}=0.04$ and then scaled to match the observed spectrum. The
    model spectra are offset by 0.3 and 0.4 vertical units for
    clarity.  }
  \label{fig:spectrum}
\end{figure*}

\subsection{Properties of star B}
The spectroscopic observation of star B shows strong absorption lines
of Na\,\textsc{D} and Ca\,\textsc{II}, while H$\alpha$ is weak. There
is some suggestion of absorption from the TiO bands near 6300 and
7000\,\AA. A comparison against templates from the library by
\citet{bbp+03} favors a K7V spectral type over K2V or M0V
templates. This classification agrees with the previous work by
\cite{srr+03}, who classified star B as a K-type dwarf.

To obtain the radial velocity and spectral parameters of star B, we
compare the observed spectrum (Fig.\,\ref{fig:spectrum}) with
synthetic spectral templates from the spectral library of
\citet{mscz05}. The templates from this library cover a range in
effective temperature $T_\mathrm{eff}$, surface gravity $\log g$,
rotational velocity $v_\mathrm{rot}\sin i$ and metallicity
$\mathrm{[M/H]}$. To account for the resolution of the observed
spectrum, we convolved the synthetic spectra with a Moffat profile
that has a width equal to the seeing and is truncated at the width of
the slit. For each combination of temperature, surface gravity and
metallicity, we used the convolved templates to find the best fitting
values for rotational velocity $v_\mathrm{rot} \sin i$ and radial
velocity $v$. These convolved, broadened and shifted spectra were then
compared with the observed spectrum while fitting the ratio of
observed and normalized flux with a second order polynomial.

To exclude possible pollution of the observed spectrum by telluric
absorption, the wavelength ranges between 5600 and 6000\,\AA\ and 8400
and 8800\,\AA\ were selected for the comparison. These ranges contain
the strong lines of the Na\,D doublet and the Ca\,\textsc{II} triplet,
but also the weak H$\alpha$ line. The observed spectrum is best
represented by the models with $\mathrm{[M/H]}=-1.0$, at an effective
temperature of $T_\mathrm{eff}=4050(50)$\,K and surface gravity $\log
g=4.28(19)$\,cgs, broadened by $v_\mathrm{rot} \sin
i=83(14)$\,km\,s$^{-1}$ at a radial velocity with respect to the
solar-system barycenter of $v=185(4)$\,km\,s$^{-1}$. The reduced
$\chi^2=3.5$ for 875 degrees-of-freedom.

\subsection{Mass, distance and reddening}
The effective temperature and surface gravity of star B are consistent
with those of a low-mass main-sequence star, as higher mass stars will
have evolved off the main-sequence and have much lower surface
gravities ($\log g<3$\,cgs). 

\begin{figure}
  \includegraphics[angle=270,width=\columnwidth]{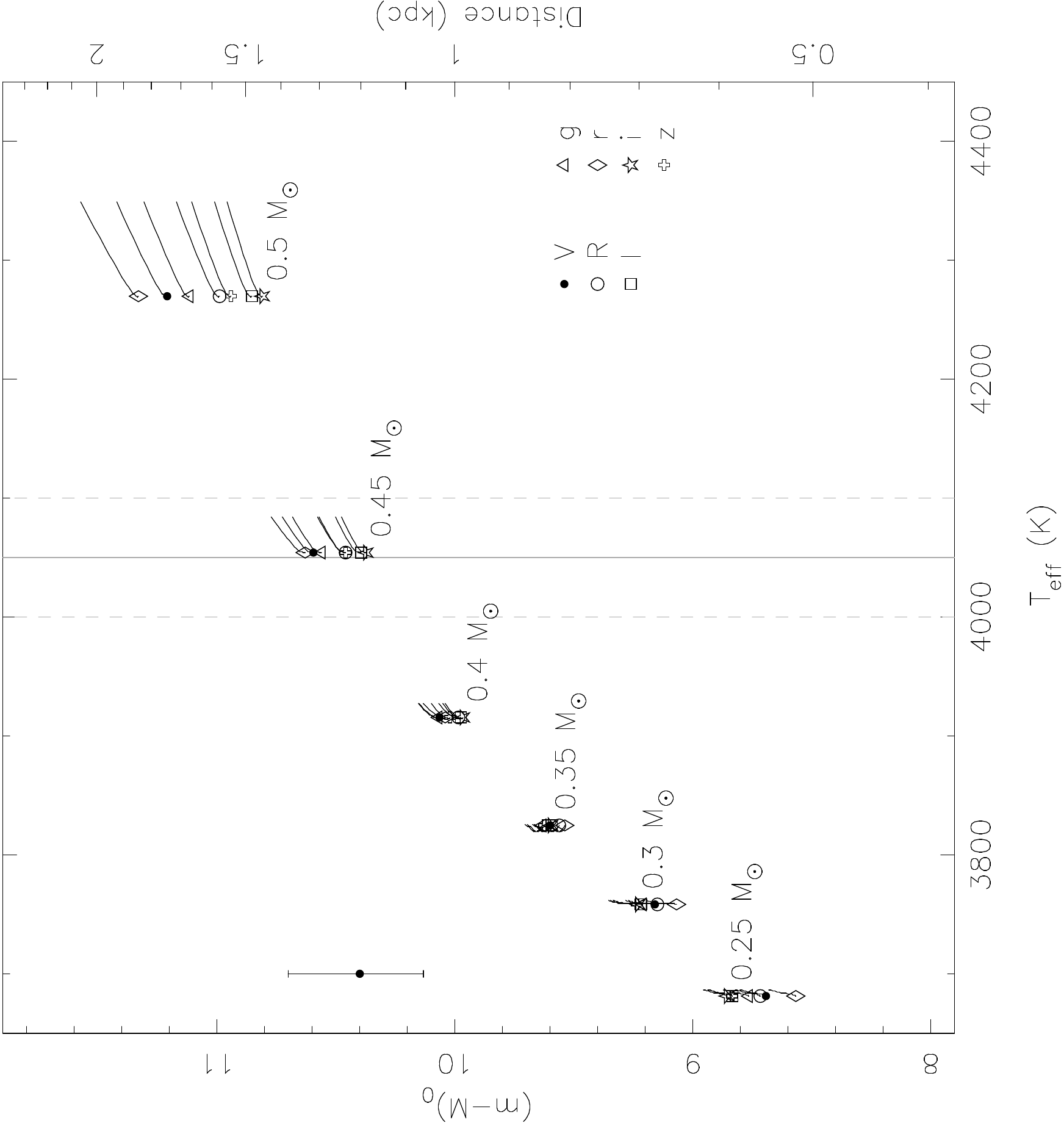}
  \caption{Mass and distance predictions from the PARSEC stellar
    evolution models by \citet{bmg+12,tbr+14,cgb+14,cbg+15} at the
    observed temperature and broadband $V\!RI$ and $griz$
    magnitudes. The observed magnitudes were corrected for reddening
    using $E_{B-V}=0.040$ from the model by \citet{gsf+14,gsf+15} and
    the \citet{sf11} extinction coefficients. The error bar on the
    left denotes the \citet{lk73} bias corrected parallax distance of
    PSR\,J1024$-$0719. }
  \label{fig:teffd}
\end{figure}

In Fig.\ref{fig:teffd} we plot the predicted effective temperatures
$T_\mathrm{eff}$ of the PARSEC stellar evolution models of
$\mathrm{[M/H]}=-1.0$ by \citet{bmg+12}, \citet{tbr+14} and
\citet{cgb+14,cbg+15} to estimate the mass and distance of star
B. Here we use our Johnson-Cousins $V\!RI$ photometry as well as the
SDSS $griz$ photometry from the ATLAS survey \citep{smc+15}, where
star B has $g=20.56(2)$, $r=19.215(10)$, $i=18.612(15)$ and
$z=18.232(19)$. We further more use the model by \citet{gsf+14,gsf+15}
to obtain the reddening $E_{B-V}$ as a function of distance towards
PSR\,J1024$-$0719. This model predicts that the reddening reaches a
maximum value of $E_{B-V}=0.040$ for distances larger than
800\,pc. Combined with the $R_V=3.1$ extinction coefficients for the
$V\!RI$ and $griz$ filters from \citet{sf11}, the observed magnitudes
are corrected for absorption and compared with the predicted absolute
magnitudes from the PARSEC models to obtain the plotted distance
modulus $(m-M)_0$ and hence distance $d$.

We find that models with masses between 0.43 and 0.46\,M$_\odot$ can
explain the observed temperature. Taking into account the scatter in
the distance moduli between the different filters, our dereddened
photometry constrains the distance to 1.1 to 1.4\,kpc. These distances
are consistent with the parallax distance (corrected for
\citealt{lk73} bias) of $d=1.20^{+0.18}_{-0.14}$\,kpc for
PSR\,J1024$-$0719 determined from pulsar timing, further confirming
the association of star B with PSR\,J1024$-$0719.

\subsection{Orbit constraints}\label{ssec:orbit}
The question now remains if star B can account for the apparent radial
acceleration and higher order spin frequency derivatives seen in the
timing of PSR\,J1024$-$0719.

Since it is unlikely that star B is in an unbound orbit around
PSR\,J1024$-$0719, we will consider a bound orbit in the following
analysis.  In Appendix\,\ref{sec:orbit}, we derive expressions for the
line-of-sight acceleration $\ddot{z}_1$, jerk $z_1^{(3)}$, snap
$z_1^{(4)}$ and crackle $z_1^{(5)}$ acting on the pulsar due to
orbital motion and also show they relate to the spin frequency
derivatives. This derivation is similar to that presented in
\citet{jr97}, though we also provide expressions for the projected
separation of the binary members on the sky, as our optical and timing
astrometry provides an additional constraint.

Besides the line-of-sight acceleration $\ddot{z}_1$, the observed
second derivative of the spin frequency $\ddot{f}$ translates into a
jerk of $z_1^{(3)}=6.1(3)\times10^{-19}$\,cm\,s$^{-3}$ along the
line-of-sight. The limits on $f^{(3)}$ and $f^{(4)}$ constrain
line-of-sight snap and crackle to
$z_1^{(4)}<4\times10^{-28}$\,cm\,s$^{-4}$ and
$z_1^{(5)}<7\times10^{-36}$\,cm\,s$^{-5}$, respectively. Furthermore,
the angular separation of $0\farcs12\pm0\farcs06$ between the pulsar
and star B at the parallax distance of $d=1.20\pm0.16$\,kpc
corresponds to a projected separation $\rho=144\pm75$\,AU.

As shown in Appendix\,\ref{sec:orbit}, six parameters are required to
describe the orbit; orbital period $P_\mathrm{b}$, semi-major axis
$a$, inclination $i$, eccentricity $e$, argument of perigee $\omega$
and true anomaly $\nu$. By treating the companion masses as known,
i.e.\ a pulsar mass of $m_1=1.5$\,M$_\odot$ and a companion mass of
$m_2=0.45$\,M$_\odot$, we can relate the orbital period $P_\mathrm{b}$
to the semi-major axis $a$, and also to the semi-major axis of the
pulsar orbit $a_1=a\ m_2/(m_1+m_2)$. To investigate which orbits are
allowed by the observations, we performed a Monte Carlo simulation
where random values for $e$, $\omega$ and $\nu$ were drawn from
uniform distributions between $0\le e<1$, $0\degr\le \omega <360\degr$
and $0\degr\le \nu<360\degr$. Furthermore, we choose random values of
$\ddot{z}_1$ from a uniform distribution between $-2\times10^{-7}$ and
$-8\times10^{-7}$\,cm\,s$^{-2}$ to account for the unknown intrinsic
spin period derivative of PSR\,J1024$-$0719. These values were used to
compute $q_2$ and $q_3$ such that the equations for $\ddot{z}_1$ and
$z_1^{(3)}$ can then be divided to obtain $a_1$ and $\sin i$. Based on
these parameters, values for $z_1^{(4)}$, $z_1^{(5)}$ and $\rho$ were
computed and compared with the observational constraints. We consider
only those orbits that yield values within $1\sigma$ of the observed
values or limits.

\begin{figure}
  \includegraphics[width=\columnwidth]{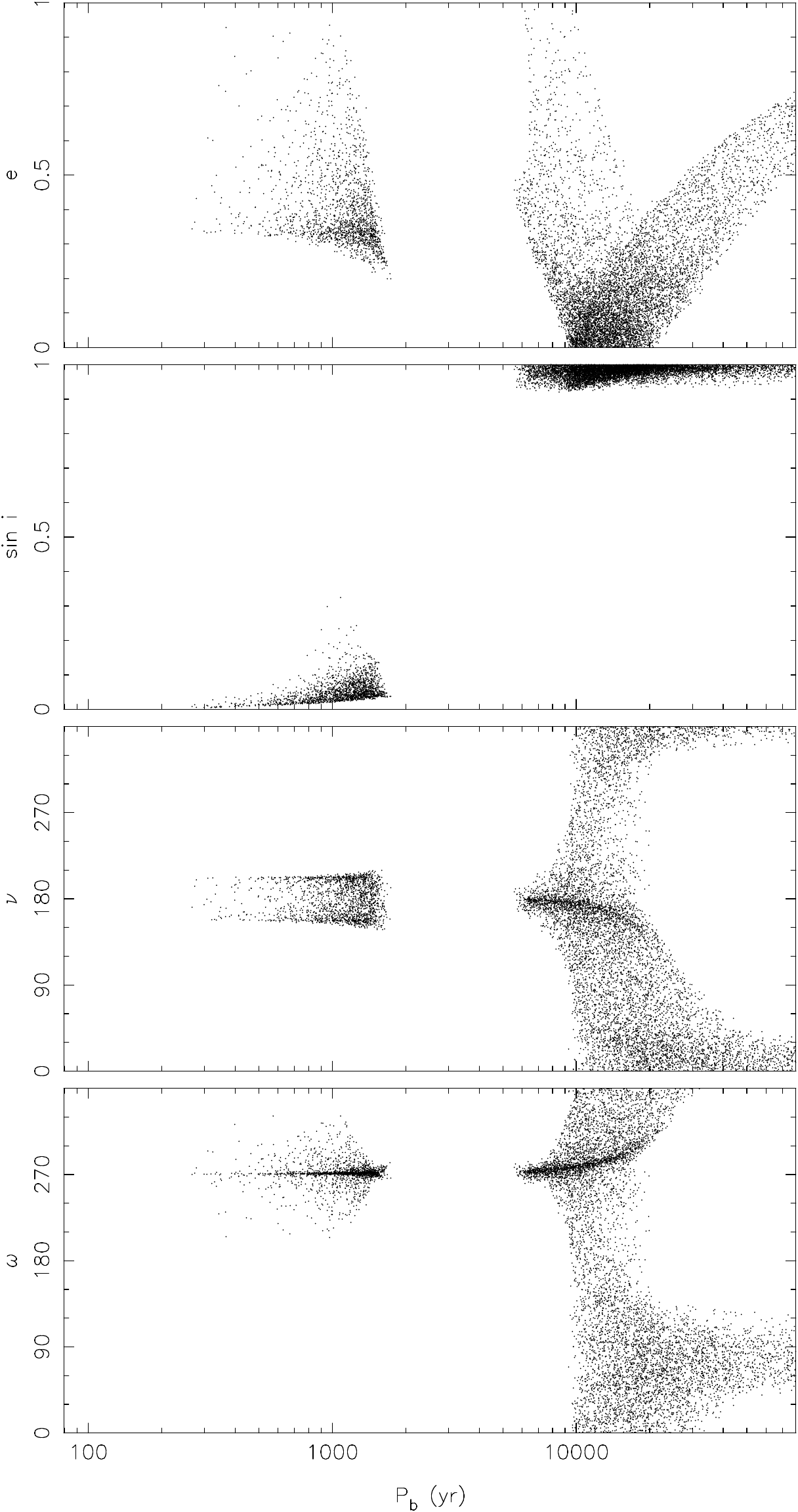}
  \caption{The results of a Monte Carlo simulation to find orbits
    which account for the observed properties of
    PSR\,J1024$-$0719. Random values for $e$, $\omega$ and $\nu$ are
    drawn from uniform distributions and used to compute
    $P_\mathrm{b}$ and $\sin i$. Only those orbits which predict
    $z_1^{(4)}$, $z_1^{(5)}$ and $\rho$ consistent with the $1\sigma$
    measurements or limits are plotted.}
  \label{fig:orbits}
\end{figure}

Figure\,\ref{fig:orbits} shows the results of the Monte Carlo
simulation. A total of 10000 possible orbits are depicted as a
function of orbital period. Two families of solutions are found. The
solutions with $\sin i<0.3$ are eccentric ($e>0.2$) and have orbital
periods up to 2000\,yr and favor $\omega\approx270\degr$ and
$\nu\approx180\degr$, i.e.\ placing the pulsar at apastron and the
apsides pointing towards the observer. The second family of solutions
is less constrained and allows for essentially all eccentricities,
also circular orbits, but requires the orbital period to be longer
than 6000\,yr and $\sin i>0.9$. Depending on the eccentricity and
orbital period, all values for $\nu$ and $\omega$ are possible.

\begin{figure*}
  \includegraphics[width=\textwidth]{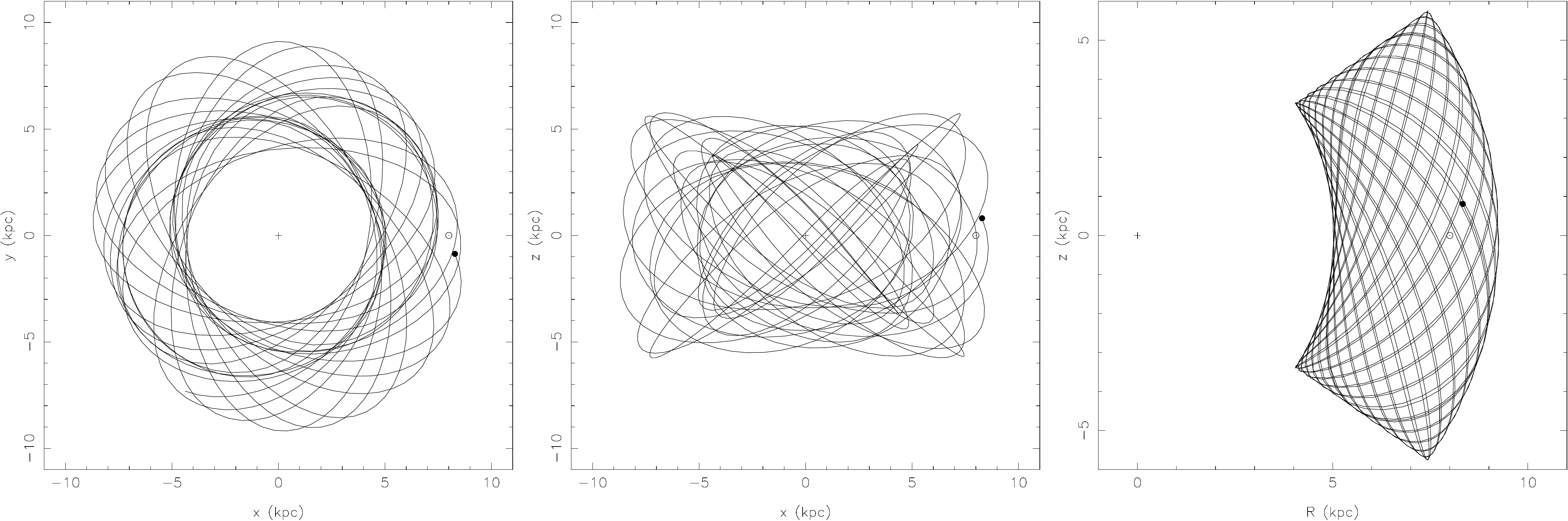}
  \caption{The present Galactic position of PSR\,J1024$-$0719 is
    indicated with the black dot, while the line traces the Galactic
    orbit of the system over the next 5\,Gyr. The location of the Sun
    at a Galactic radius of $R=8$\,kpc is denoted with $\odot$, while
    the Galactic center is at the plus sign. It is clear that
    PSR\,J1024$-$0719 is in an eccentric and inclined orbit in the
    Galaxy.}
  \label{fig:galorbit}
\end{figure*}

\subsection{Space velocity}\label{ssec:velocity}
The radial velocity measurement of star B of
$V_\mathrm{R}=185\pm4$\,km\,s$^{-1}$ complements the transverse
velocity $V_\mathrm{T}=341\pm45$\,km\,s$^{-1}$ from the proper motion
measurement of PSR\,J1024$-$0719 and allows us to determine the 3D
velocity of the system in the Galaxy. Based on the orbital solutions
presented in \S\,\ref{ssec:orbit}, we find that the corrections on the
transverse velocity of PSR\,J1024$-$0719 and the radial velocity of
star B with respect to the binary center of mass are negligible in
comparison with the uncertainties on $V_\mathrm{T}$ and
$V_\mathrm{R}$. As such, we treat $V_\mathrm{T}$ and $V_\mathrm{R}$ as
representative of the velocity of the binary center of mass, yielding
a total velocity of $388\pm39$\,km\,s$^{-1}$ with respect to the
solar-system barycenter. Converting these velocities into the Galactic
frame using the algorithm by \citet{js87}\footnote{With a rotation
  matrix for the J2000 input frame.} and correcting for solar motion
with the values from \citet{hbr+05}, the resulting peculiar motion
with respect to the LSR is $(U,V, W)=(-77, -353,
-129)\pm(3,30,34)$\,km\,s$^{-1}$.

The distance, location and proper motion of PSR\,J1024$-$0719 place it
0.82\,kpc above and moving towards the Galactic plane. Using the
\texttt{MWPotential2014} Galactic potential and the orbit integrator
implemented in the \textsc{Galpy} software package by \citet{bov15},
we numerically integrate the orbit of PSR\,J1024$-$0719 in the
Galaxy. The result of the integration is plotted in
Fig.\,\ref{fig:galorbit}, which shows that the Galactic orbit of
PSR\,J1024$-$0719 is eccentric and inclined. The Galactic orbit varies
between Galactic radii of 4 and 9\,kpc ($e\sim0.5$) and reaches up to
5.8\,kpc above the Galactic plane. The vertical velocity of the system
as it crosses the Galactic plane varies between between 120 and
220\,km\,s$^{-1}$.  With these properties, PSR\,J1024$-$0719 can be
classified as belonging to the Galactic halo \citep{bg16}.

\section{A triple star formation scenario}\label{sec:formation}
The standard formation scenario for millisecond pulsars is not
applicable in the case of PSR\,J1024$-$0719. In that scenario, a
period of mass and angular momentum transfer from a stellar binary
companion spins up an old neutron star \citep{acrs82,bh91}. The
outcome of that evolutionary channel is a ``recycled'' radio
millisecond pulsar orbiting the white dwarf remnant of the
main-sequence star in a short ($0.1<P_\mathrm{b}<200$\,d) and
circular orbit (e.g.\ \citealt{tau11,tlk12}). This channel has been
confirmed observationally by the identification of several white dwarf
binary companions to MSPs
(e.g.\ \citealt{kbjj05,bkkv06,akk+12,kbk+13}).

In the case of PSR\,J1024$-$0719, the current binary companion is a
main-sequence star and thus can not have recycled the
pulsar. Currently we know of only one other MSP with a main-sequence
companion; PSR\,J1903+0327, which is a 2.15\,ms period MSP in a
95\,day eccentric ($e=0.44$) orbit \citep{crl+08} around a
$1.0$\,M$_\odot$ G-dwarf companion \citep{fbw+11,ksf+12}. This system
is believed to have evolved from a hierarchical triple consisting of a
compact low-mass X-ray binary, accompanied by the G dwarf in a wide
orbit. The inner companion then was either evaporated by the pulsar
\citep{fbw+11}, or the widening of the orbit of the X-ray binary due
to mass transfer led to a dynamical instability which ejected the
inner companion from the triple \citep{fbw+11,phln11}.

For PSR\,J1024$-$0719, we argue below that the ejection scenario is
unlikely based on its orbital constraints. The evaporation scenario
proposed in \citet{fbw+11} is only described qualitatively. To obtain
a quantitative description, we investigate if PSR\,J1024$-$0719 could
have evolved from a triple system in which the inner companion was
evaporated. Further considerations of mass transfer, stability, and
the effects of supernovae in triple systems can be found in
\citet{tv14}.

\subsection{Summary of our model}\label{ssec:triple_summary}
There are a number of criteria which must be fulfilled in order to
produce a stable triple system which would eventually leave behind
PSR\,J1024$-$0719: i) the triple system must remain bound after the
supernova (SN) explosion which creates the neutron star (NS), ii) to
survive the SN, the secondary and the tertiary star must therefore be
brought close to the exploding star via significant orbital angular
momentum loss, e.g.\ in a common envelope, iii) the SN explosion must
result in a large 3D systemic velocity and cause the tertiary star to
be almost (but not quite) ejected, in order to explain the observed
properties of PSR\,J1024$-$0719, iv) the post-SN triple system must
remain dynamically stable on a long timescale (i.e.\ avoid chaotic
three-body interactions) and v) the inner binary must evolve to
evaporate the secondary star after the LMXB recycling phase, leaving
behind PSR\,J1024$-$0719.

\begin{figure}
  \begin{center}
    \includegraphics[width=\columnwidth,angle=0]{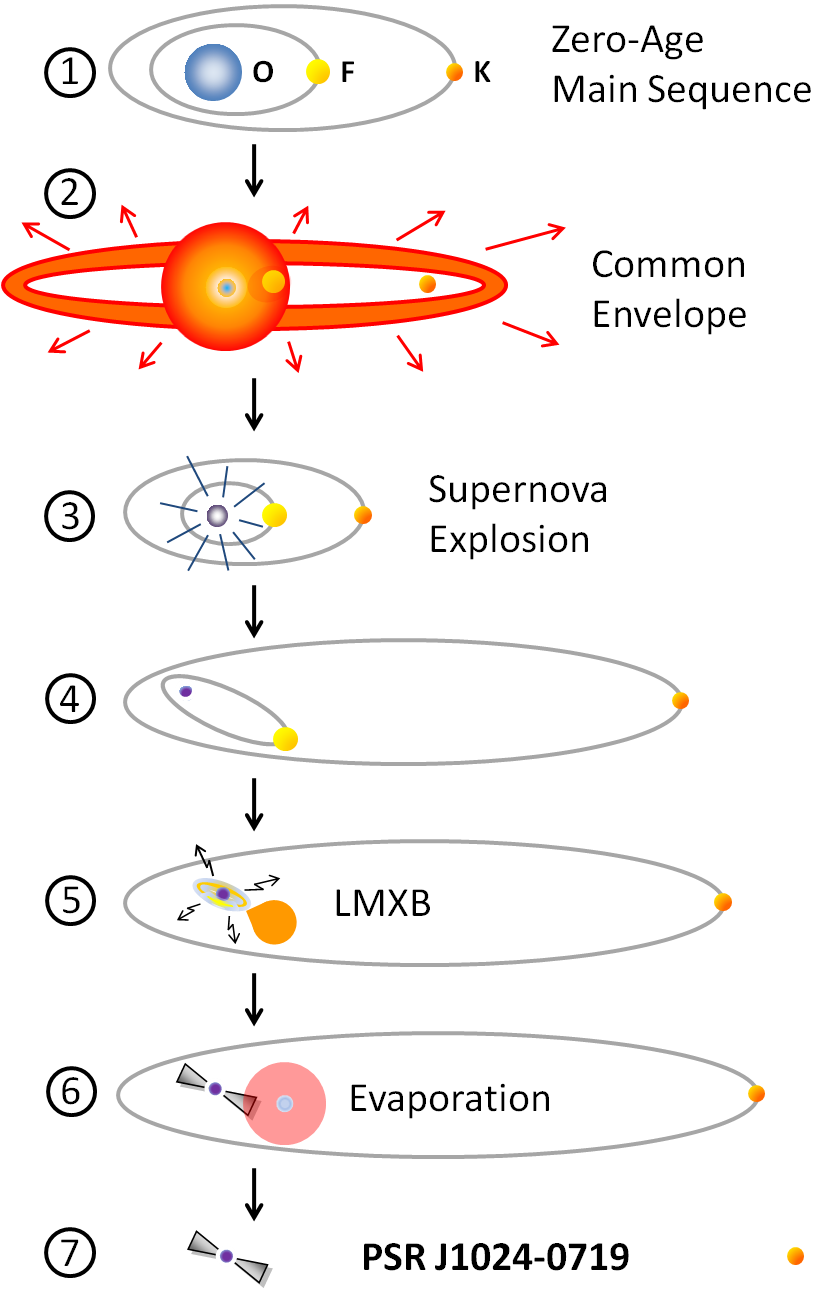}
    \caption{Illustration of the formation of PSR~J1024$-$0719 via a
      triple star scenario: 1) A triple system is formed with a
      massive star (an O-star, the progenitor of the neutron star,
      NS), an F-star (needed to recycle the NS) and a K-dwarf (star B
      as presently observed).  2) When the massive star becomes a
      giant it expands and engulfs its two orbiting stars.  3) The
      supernova (SN) explosion.  4) The dynamically stable post-SN
      system.  5) Mass-transfer and recycling of the NS during the
      low-mass X-ray binary (LMXB) phase of the inner binary.  6)
      Evaporation of the remnant of the secondary star from the
      energetic wind of the recycled millisecond pulsar.  7) The
      present day system PSR\,J1024$-$0719.  See text for a discussion
      of the various stages.}
  \label{fig:cartoon}
  \end{center}
\end{figure}

We find that the following model is able to account for the criteria
listed above, albeit with significant fine-tuning during the SN
explosion. The model is illustrated in
Fig.\,\ref{fig:cartoon}. Starting on the zero-age main-sequence
(ZAMS), the system consists of a roughly 16 to 18\,M$_\odot$ primary
O-star and two low-mass companions: a secondary F-star with a mass of
$M_2=1.1$ to 1.5\,M$_\odot$ and a tertiary K-dwarf with a mass
$M_3\simeq 0.45$\,M$_\odot$, the presently observed star B
(stage~1). After a common envelope phase (stage~2), where the extended
envelope of the primary engulfed the other two stars prior to its
ejection, the orbital period of the inner binary, $P_\mathrm{b,12}$
was reduced to a few days and the orbital period of the tertiary star
was $P_\mathrm{b,3}\approx 100$\,d. Following a phase of wind-mass
loss\footnote{Given the low metallicity of star B, we only expect a
  moderate amount of wind-mass loss from the helium star prior to its
  collapse.} from the exposed (initially) $\sim\!6$\,M$_\odot$ helium
core, that star reaches a mass of $M_1\simeq 4.3$\,M$_\odot$ (and
$P_\mathrm{b,12}=3.0$\,d) when it collapses and undergoes a SN
explosion (stage~3).  Whereas the inner binary remains compact (but
now with a fairly high eccentricity), the dynamical effects of the SN
causes the tertiary star to be almost (but not quite) unbound,
reaching an orbital period of $10^2$ to $10^3$\,yr in an extremely
eccentric orbit ($e\simeq 1$), and yet remaining in a long-term
dynamically stable three-body system (stage~4, see
Section\,\ref{ssec:stability}).  A combination of tidal damping and
magnetic braking eventually causes the secondary star to fill its
Roche-lobe while still on the main sequence and initiate mass transfer
to the NS, i.e.\ the inner binary becomes an LMXB system (stage~5).
The LMXB system is converging, i.e.\ evolving to a small orbital
period, and the secondary star may either form a low-mass helium white
dwarf companion with a mass of $0.16-0.20$\,M$_\odot$ \citep{itl14} or
continue evolving through the period minimum and become a
semi-degenerate ''black widow"-type companion with a mass of a few
$0.01$\,M$_\odot$ \citep{prp02,ccth13}.  Meanwhile, the accreting NS
is recycled and turns on as an energetic millisecond pulsar which will
evaporate its companion star completely (stage~6), thereby leaving
behind a system with the presently observed properties of
PSR\,J1024$-$0719 (stage~7).

\subsection{The dynamical effects of the SN explosion}\label{ssec:SN}
We now describe in more detail the dynamically important properties of
our model. The dynamical effects of asymmetric SNe in binaries,
leading to either bound or disrupted systems, has been studied in full
detail \citep{fv75,hil83,tt98}.  A general discussion of dynamical
effects of asymmetric SNe in hierarchical multiple star systems is
found in \citet{pcp12}, and references therein.  Here, we follow their
method and apply a two-step process, where we first calculate the
effects of an asymmetric SN explosion in the inner binary and then
treat the inner binary as an effective point mass with respect to the
tertiary star and calculate the dynamical effects of the SN on the
outer binary. Hence, we use the obtained recoil velocity of the inner
binary as a ``kick'' velocity added to this effective point-mass star
representing the inner binary and then calculate the solution for the
outer binary.  We assume the SN explosion is instantaneous and for
simplicity we neglect in our example any dependence on the orbital
phase or inclination of the inner binary with respect to the outer
binary.

Assuming a circular pre-SN orbit, the change of the binary semi-major
axis as a result of an asymmetric SN, is given by \citep{hil83}:
\begin{equation}
  \frac{a}{a_0}=\left[ \frac{1-(\Delta M/M_0)}{1-2(\Delta
      M/M_0)-(w/v_c)^2-2\cos\theta\;(w/v_c)} \right] \;,
  \label{eq:kick}
\end{equation}
where $a_0$ is the pre-SN semi-major axis (radius), $a$ the post-SN
semi-major axis, $\Delta M$ the effective mass loss during the SN
(here in our example, we shall assume 2.9\,M$_\odot$ to be lost when
the $M_1=4.3$\,$M_\odot$ exploding star leaves behind a NS with a
gravitational mass of $M_\mathrm{NS}=1.4$\,M$_\odot$), $M_0=M_1+M_2$
the pre-SN total mass of the binary, $v_c=\sqrt{GM_0/a_0}$ the pre-SN
orbital velocity of the exploding star in a reference fixed on the
secondary companion star, $w$ the magnitude of the kick velocity, and
$\theta$ the angle ($0\degr-180\degr$) between the kick velocity
vector, $\vec{w}$ and the pre-SN orbital velocity vector, $\vec{v}_c$.
When calculating the dynamical effects for the outer binary, we simply
use Eq.~(\ref{eq:kick}) with $M_1'=M_1+M_2$ and $M_2'=M_3$.

In the case of a purely symmetric SN ($w=0$), a binary will always
disrupt if $\Delta M/M_0 > 0.5$. This follows from the virial theorem
and is also seen when the denominator in Eq.~(\ref{eq:kick}) becomes
negative).  Since the tertiary star (star B) is almost ejected
from the system, we can estimate a critical mass of the exploding star
to be 4.55\,M$_\odot$ (for which $\Delta M/M_0 = 0.5$, if we assume
$M_2=1.3$\,M$_\odot$). In our selected model, we then chose a slightly
smaller mass of $M_1=4.3$\,M$_\odot$ since we also need to apply a
kick to obtain a large systemic velocity.

From Eq.~(\ref{eq:kick}), we can find a critical value for the kick
angle, $\theta _\mathrm{crit}$ such that a binary will always dissociate
if $\theta<\theta_\mathrm{crit}$. Hence, by integrating over all kick
angles, assuming an isotropic distribution of kick directions, we can
easily calculate the probability $P(\theta>\theta_\mathrm{crit}$) for the
binary to remain bound \citep{hil83}:
\begin{eqnarray}
  \label{eq:prob}
  P_{\rm bound} & = & \int _{\theta_\mathrm{crit}}^{180^{\circ}}
  \frac{1}{2}\,\sin\theta d\theta=\frac{1}{2}(1+\cos\theta_\mathrm{crit})
  \\ \nonumber & = & \frac{1}{2}\left\{ 1+\left[ \frac{1-2(\Delta
      M/M_0)-(w/v_c)^2}{2\,(w/v_c)} \right]\right\}\;.
\end{eqnarray}

\begin{figure}
  \centering
  \includegraphics[angle=270,width=\columnwidth]{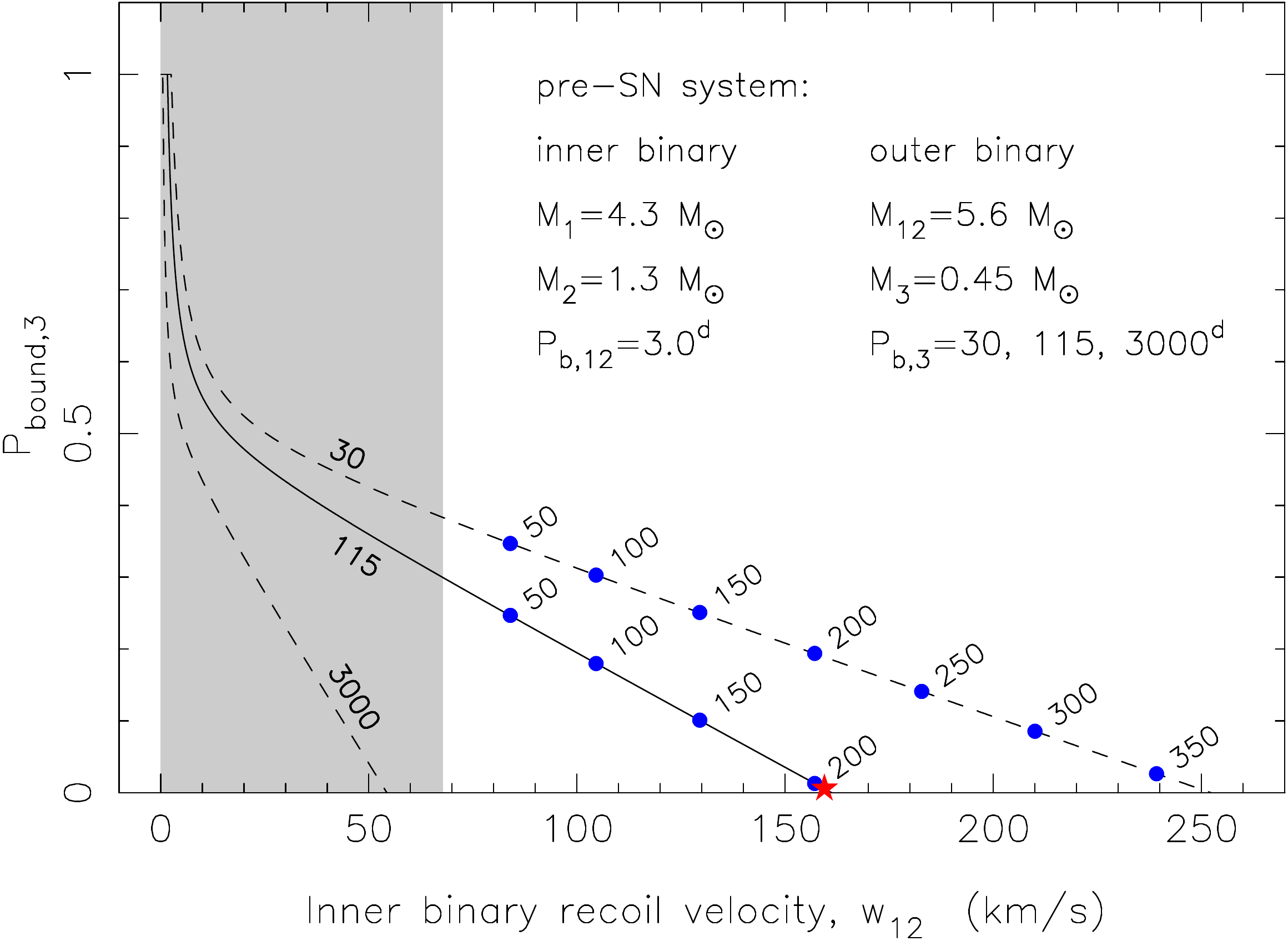}
  \caption{Probability for a triple system to survive a given recoil
    velocity, $w_\mathrm{12}$ obtained by the inner binary due to a
    SN.  The three lines are for a pre-SN orbital period of the
    tertiary star of $P_\mathrm{b,3}=30$, 115 and 3000\,d. The bullet
    points represent average values of $w_\mathrm{12}$ for the stated
    kick magnitudes (in km\,s$^{-1}$) imparted onto the newborn
    1.4\,M$_\odot$ NS in the inner binary, calculated from Monte Carlo
    simulation of $10^7$ SNe and using an isotropic distribution of
    kick directions for both $w$ and $w_{12}$.  When calculating
    $\langle w_\mathrm{12} \rangle$, only triple systems for which the
    inner binary avoided merging were considered.  However, the
    long-term stability of the surviving triple systems is not
    included here. The assumed pre-SN mass of the exploding star is
    $M_1=4.3$\,M$_\odot$, the secondary star has a mass of
    $M_2=1.3$\,M$_\odot$ and the tertiary star has a mass of
    $M_3=0.45$\,M$_\odot$. The pre-SN orbital period of the inner
    binary is $P_\mathrm{b,12}=3.0$\,d. The red star at
    $w_{12}=159$\,km\,s$^{-1}$ ($P_\mathrm{b,3}=115$\,d) marks the
    example investigated further in this scenario. }
\label{fig:prob}
\end{figure}

\begin{figure}
  \centering
  \includegraphics[angle=270,width=\columnwidth]{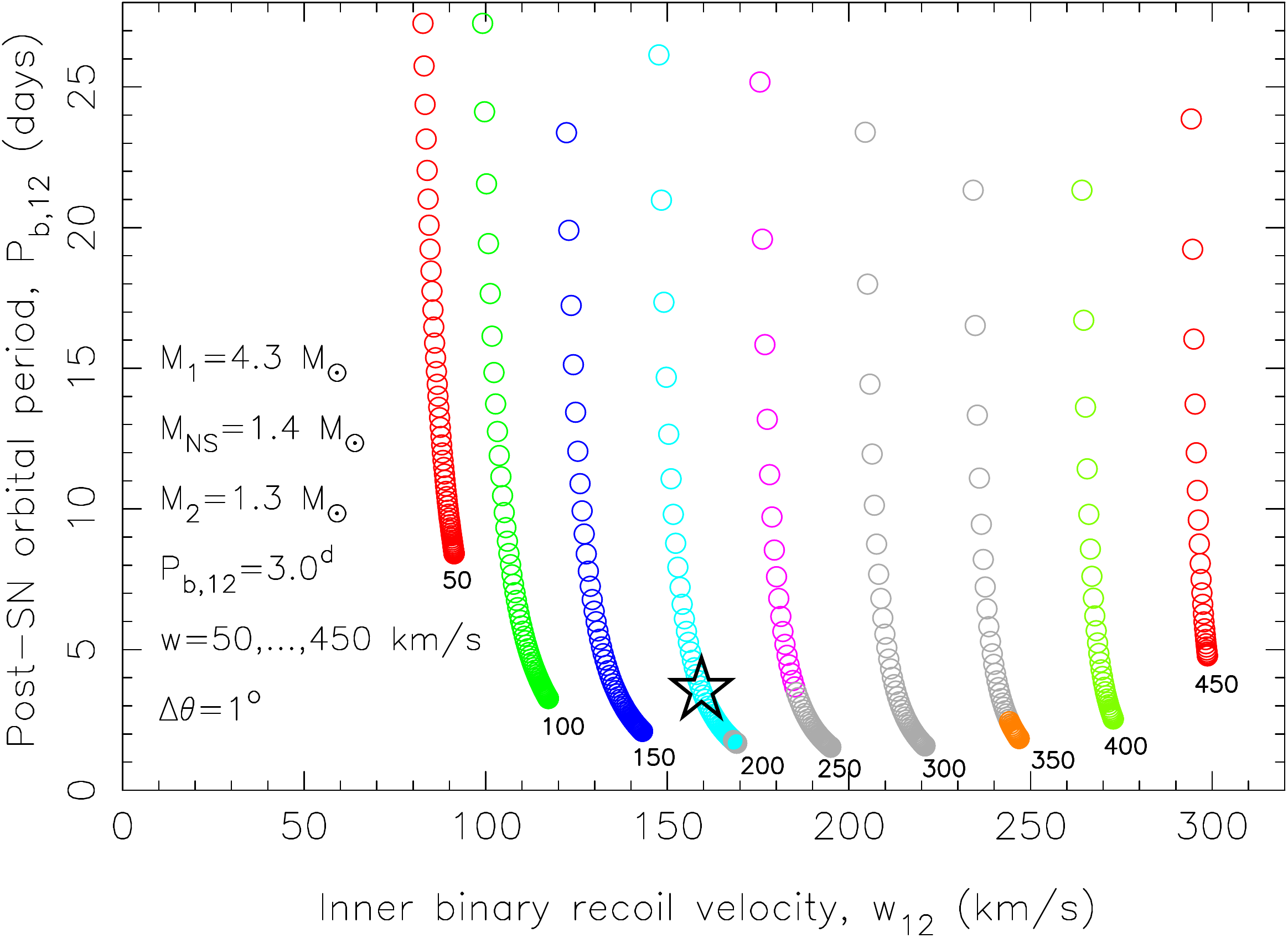}
  \caption{ Post-SN orbital period of the inner binary
    ($P_\mathrm{b,12}$) as a function of its resulting recoil velocity
    ($w_{12}$) and kick velocity ($w$) imparted onto the NS. Each
    colored track is for a given value of $w$ (between 50 and
    450\,km\,s$^{-1}$) and plotted along the tracks are solutions for
    different values of the kick angle $\theta$ (in steps of
    $1\degr$).  Grey circles represent systems that merge during the
    SN. The open star indicates the selected example which we
    investigate in more detail.  }
\label{fig:SN_inner}
\end{figure}

\begin{figure}
  \centering
  \includegraphics[angle=270,width=\columnwidth]{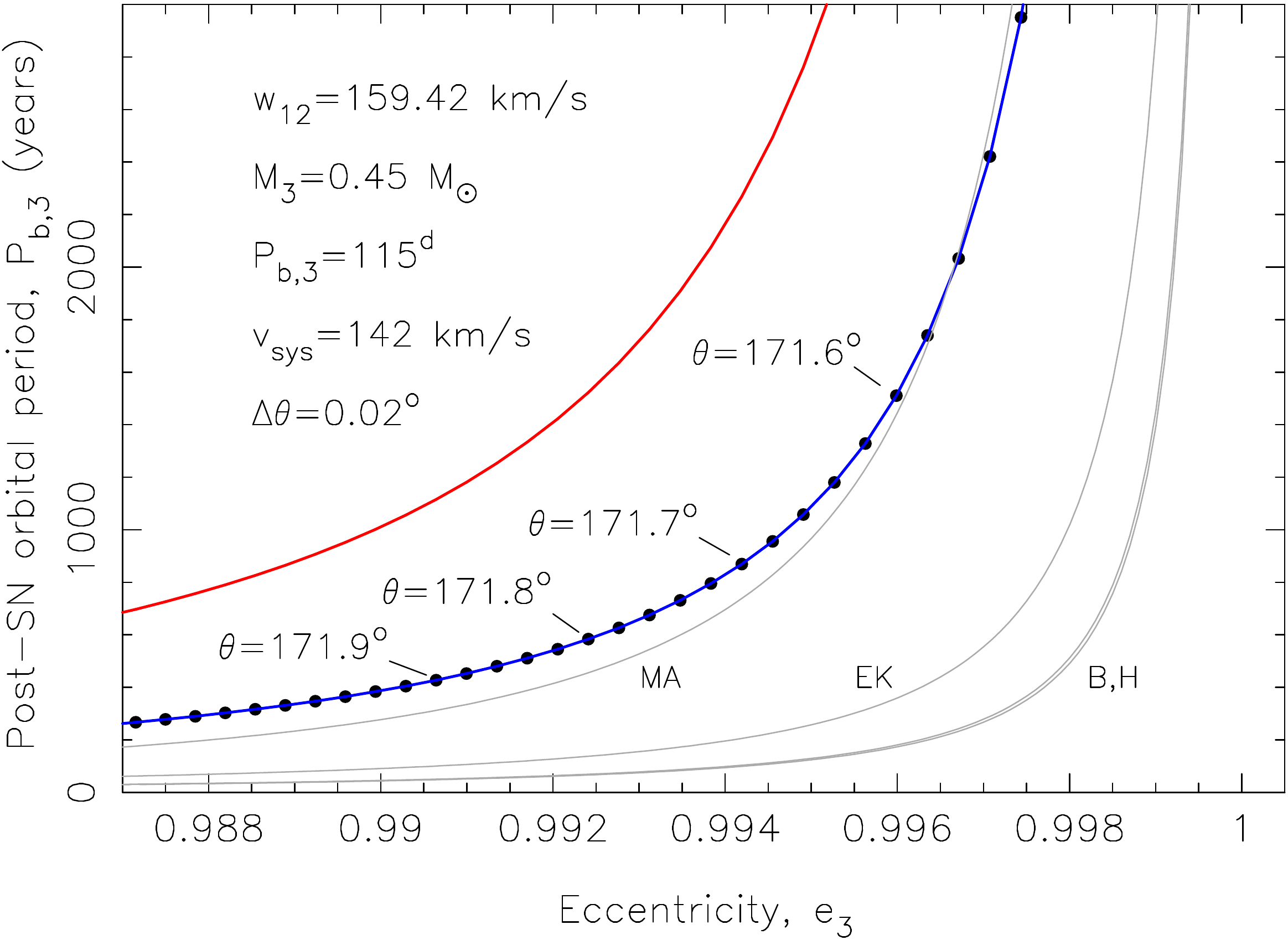}
  \caption{Post-SN orbital period of the tertiary star,
    $P_\mathrm{b,3}$ as a function of the eccentricity of its orbit
    using our selected inner binary model from
    Fig.~\ref{fig:SN_inner}. The pre-SN model has an orbital period of
    $P_\mathrm{b,3}=115$\,d. Solutions are shown along the blue curve
    for different values of the kick angle, $\theta$ in steps of
    $0.02\degr$. The grey lines shown minimum orbital periods for a
    given eccentricity to ensure a long-term stable three-body
    configuration. The red line shows the expected orbital period
    after evaporation of the secondary star -- see text for
    discussions. }
  \label{fig:SN_outer}
\end{figure}

In Fig.~\ref{fig:prob}, we plot examples of the
probability\footnote{The plotted probabilities do not take into
  account the specific requirements on the value of the post-SN
  orbital period of the tertiary star necessary for forming
  PSR\,J1024$-$0719. If including this specific constraint, the
  probabilities shown would be much lower.  Similarly, for estimating
  the resulting values of $w_{12}$ we did not filter out those
  surviving triples which would not survive the SN in long-term stable
  orbits (see Section\,\ref{ssec:stability}).}  of the outer binary to
remain bound as a function of the resulting recoil velocity, $w_{12}$
of the inner binary, for three different pre-SN values of the orbital
period of the tertiary star ($P_\mathrm{b,3}=30$, 115 and
3000\,days). The dots indicate average values of $w_{12}$ for a given
kick velocity added to the newborn NS (between 50 and
350\,km\,s$^{-1}$). To calculate each of these values, we simulated
$10^7$ SN explosions using Monte Carlo techniques. As a result of mass
loss of the inner binary, the minimum value of $w_{12}$ for any bound
solution of the triple system is 67.8\,km\,s$^{-1}$, thus excluding
any solutions in the grey-shaded (forbidden) region in
Fig.~\ref{fig:prob}.  It is clearly seen that the probability of
remaining bound is small for large values of $P_\mathrm{b,3}$
(e.g.\ there are no possible solutions if $P_\mathrm{b,3}=3000$\,d).
The star at $w_{12}=159$\,km\,s$^{-1}$ is for the selected example
investigated further in this scenario (Figs.\,\ref{fig:SN_inner} and
\ref{fig:SN_outer}).

The values of the pre-SN orbital periods of the tertiary star
($P_\mathrm{b,3}$) are limited to an interval where it cannot be too
large (otherwise the system would dissociate) or too small (in which
case the pre-SN triple system would not be stable due to three-body
interactions between the stars. Besides reproducing the orbital period
of the K-dwarf (star B) in PSR\,J1024$-$0719, we must also make sure
to reproduce the large systemic velocity, $v_\mathrm{sys}$ of
PSR\,J1024$-$0719.  As derived in Section\,\ref{ssec:velocity}, the
vertical velocity of PSR\,J1024$-$0719 when it crosses the Galactic
plane must be between 120 and 220\,km\,s$^{-1}$. Hence, our model must
produce {\it at least} $v_\mathrm{sys}=120$\,km\,s$^{-1}$ for an
optimal orientation of $v_\mathrm{sys}$. A large value of
$v_\mathrm{sys}$ requires a large value of $w_{12}$, which depends on
the magnitude of the kick, $w$ imparted onto the newborn NS during the
SN as well as the value of the pre-SN orbital period of the tertiary
star $P_\mathrm{b,3}$.

In Fig.~\ref{fig:SN_inner} we demonstrate how $w_{12}$ increases with
the kick velocity, $w$.  For these calculations, we assumed in all
cases a pre-SN inner binary with circular orbit and
$M_1=4.3$\,M$_\odot$, $M_\mathrm{NS}=1.4$\,M$_\odot$,
$M_2=1.3$\,M$_\odot$, $P_\mathrm{b,12}=3.0$\,d and varied $w$ between
50 and 450\,km\,s$^{-1}$ ($w=0$ is not included as it would lead to
dissociation of the system). Each point (colored circle) represents a
value of the kick angle, $\theta$ (in steps of 1\degr) for a given
value of $w$. The magnitude of $w$ is given at the bottom of each
track which ends at $\theta = 180\degr$. For clarity, we assumed here
that the kick was in the pre-SN plane of the inner binary -- i.e.\ the
second kick angle $\phi=0\degr$ or $\phi=180\degr$, where $\phi$ is
the angle between the projection of the kick velocity vector onto a
plane perpendicular to the pre-SN velocity vector of the exploding
helium star and the pre-SN orbital plane \citep[e.g. Fig.~1
  in][]{tt98}.  The post-SN orbital period of the inner binary,
$P_\mathrm{b,12}$ is seen to decrease monotonically with larger values
of $\theta$, i.e.\ when the kick is directed backwards compared to the
pre-SN orbital motion of the exploding star.  Similarly, we have that
$P_\mathrm{b,12}\rightarrow \infty$ for $\theta \rightarrow
\theta_\mathrm{crit}$ (cf. Eq.~\ref{eq:kick}). Grey colors mark
systems which immediately merge after the SN, i.e.\ systems where the
periastron separation of the post-SN inner binary is smaller than the
radius ($\sim\!1.1$\,R$_\odot$) of the secondary star (here assuming a
$1.3$\,M$_\odot$ main-sequence F-star). The secondary star in our
model must have a mass between 1.1 and 1.5\,M$_\odot$ to be massive
enough to evolve within a Hubble time and still be light enough to
possess a convective envelope, which is necessary for magnetic braking
to operate during the LMXB phase.

The open black star symbol in Fig.~\ref{fig:SN_inner} indicates our
selected example model (see below) for which $w=200$\,km\,s$^{-1}$ and
$\theta=140\degr$, leading to post-SN values of $P_\mathrm{b,12}=3.54$\,d,
$e_{12}=0.768$ and $w_{12}=159.42$\,km\,s$^{-1}$. The surviving systems for
which $w=350-450$\,km\,s$^{-1}$ produce too large post-SN
eccentricities to ensure a long-term stable triple system.

The post-SN eccentricity is given by:
\begin{equation}
  e = \sqrt{1+\frac{2E_{\rm orb}J_{\rm orb}^2}{\mu\,G^2M_{\rm NS}^2M_2^2}} \;,
  \label{eq:ecc}
\end{equation}
where the post-SN orbital energy of the system is given by: $E_{\rm
  orb}=-GM_{\rm NS}M_2/2a$, and the orbital angular momentum is given
by:
\begin{equation}
  J_{\rm orb} =
  a_0\,\mu\sqrt{(v_c+w\cos\theta)^2+(w\sin\theta\sin\phi)^2} \;,
  \label{eq:J_orb}
\end{equation}
where $\mu$ is the reduced mass of the post-SN binary. We neglected
any shell impact effects on the companion star since even for SN~Ib/c
where a significant shell is ejected, the impact effect on the orbital
dynamics is small if the pre-SN separation is larger than a few
$R_\odot$ \citep{ltr+15}.

As we have seen already in Fig.~\ref{fig:prob}, a pre-SN orbital
period of the tertiary star of $P_\mathrm{b,3}=3000$\,d
($\sim\!8$\,yr) is much too wide to keep the system bound after the SN
explosion. Therefore, the value of $P_\mathrm{b,3}$ must have been
much smaller and in our chosen example $P_\mathrm{b,3}=115$\,d. As a
result, the eccentricity of the post-SN system must become very large
in order for the K-dwarf (star B) to reach post-SN orbital periods of
$P_\mathrm{b,3}=200-1400$\,yr, as derived in Section~\ref{ssec:orbit}.

In Fig.~\ref{fig:SN_outer}, we plot the post-SN orbital period of the
tertiary star, $P_\mathrm{b,3}$ as a function of the eccentricity of
its orbit using our selected inner binary model with
$w=200$\,km\,s$^{-1}$ and $\theta=140\degr$, leading to
$w_{12}=159$\,km\,s$^{-1}$, $P_\mathrm{b,12}=3.54$\,d and $e_{12}=0.768$
for the post-SN inner binary, and assuming a pre-SN orbital period of
the tertiary star of $P_\mathrm{b,3}=115$\,d. We only find solutions
for a system mimicking PSR\,J1024$-$0719 for a very small interval of
$171.6\degr<\theta<172.0\degr$ given our particular choice of pre-SN
parameters.  The values of $\theta$ are, as expected, close to the
critical value of $\theta_\mathrm{crit}=171.4\degr$ where the system
dissociates ($e_3>1$). We can calculate $\theta_\mathrm{crit}$ for
this system from Eq.~(\ref{eq:prob}). The red line indicates the final
orbital period of the tertiary star as a consequence of orbital
widening when the secondary star has been evaporated (assuming here
that the NS has accreted 0.1\,M$_\odot$ during the LMXB phase).  The
final 3D systemic velocity of our model is
$v_\mathrm{sys}=142$\,km\,s$^{-1}$, which is sufficient to explain the
Galactic motion of PSR\,J1024$-$0719 (Fig.\,\ref{fig:galorbit}). As we
discuss below, the post-SN orbit of the tertiary star (star~B, the
K-dwarf) cannot be too eccentric since this would prevent the post-SN
triple system from remaining bound on a long timescale.

\subsection{Long-term stability of a triple system}\label{ssec:stability}
For all our solutions of bound post-SN triples, we must check if their
mutual stellar orbits are dynamically stable on a long-term timescale.
One must keep in mind that it is expected to last several Gyr from
the SN explosion until the secondary star evolves through the LMXB
phase and finally becomes evaporated. In all this interval of time,
the triple system must be stable.  This means that for a given
eccentricity of the outer binary, we must require that the ratio of
the periastron separation of the tertiary star to the semi-major axis
of the inner binary, ($R_\mathrm{peri}/a_{12}$) be larger than a
critical value.  Otherwise, three-body interactions will cause the
hierarchical triple to be unstable and eject of one of the stars via
tidal disruption or chaotic energy exchange \citep{ma01}.  This
criterion translates into the requirement that the post-SN orbital
period of the tertiary star, $P_\mathrm{b,3}$, with a given
eccentricity, must be larger than a critical value.

A number of stability criteria for triple systems have been proposed
over the last four decades \citep[see][for an overview]{mik08}.  The
grey lines in Fig.~\ref{fig:SN_outer} show stability criteria for
triple systems as suggested by \citet{ma01,ek95,bai87,har72},
indicated by MA, EK, B and H, respectively. The earlier of these
studies have been derived empirically for restricted configurations.
The most stringent of the stability criteria is by far that of
\citet[Eq.\,90]{ma01};
\begin{equation}
  \left( \frac{R_\mathrm{peri}}{a_{12}} \right) _\mathrm{crit}=2.8
  \left [ \left (1+\frac{M_3}{M_1+M_2}\right )\frac{1+e_{3}}{\sqrt{1-e_{3}}} \right ]^{2/5},
\end{equation}
where $R_\mathrm{peri}=a_3(1-e_3)$, and which can be combined with
Kepler's third law to yield critical orbital periods as plotted in
Fig.\,\ref{fig:SN_outer}.  We note that for our chosen example, the
triple system would become unbound for $P_\mathrm{b,3}\ga 2000$\,yr.
Nevertheless, for $P_\mathrm{b,3}<2000$\,yr (corresponding to a
present binary orbital period of $\la 5000$\,yr after evaporation of
the secondary star) we can reproduce PSR\,J1024$-$0719 for the triple
model investigated here.

Finally, it should be noted that the strict stability criteria of
\citet{ma01} was derived for coplanar prograde motion in accordance
with a recoil velocity of the inner binary in the orbital plane of the
outer orbit.  Hence, this criteria represents an upper limit for
stability of non-coplanar systems since triple systems whose inner and
outer orbits are inclined, despite Kozai interactions \citep{koz62},
will be more stable than coplanar prograde systems with the same mass
ratios and eccentricities. \citet{ma01} estimate that for non-coplanar
systems, the ratio ($R_\mathrm{peri}/a_{12}$) could be around
30~per~cent smaller and still result in long-term stability against
escape in the three-body system. This would slightly enhance the
probability for producing a system like PSR\,J1024$-$0719.

We have demonstrated that PSR\,J1024$-$0719 could have formed via a
triple system in the Galactic disk. However, estimating the
probability for this formation channel would require an investigation
combining population synthesis, gravitational dynamics, detailed
stellar evolution and hydrodynamical interactions of the common
envelope phase, which is much beyond the scope of this paper. While it
is difficult to estimate the rate at which binaries like
PSR\,J1024$-$0719 could be produced from a triple star scenario, we
have demonstrated that this scenario {\it is} possible, although
severe fine-tuning is required for this to happen.  Nevertheless, so
far we have only detected one such system in our Galaxy.

\subsection{Alternative models}\label{ssec:alternative}
A different formation scenario has been suggested by
\citet{kkn+16}. In the scenario, PSR\,J1024$-$0719 has been recycled
in the center of a globular cluster, where an exchange encounter with
another cluster star or binary led to the ejection of the pulsar from
the core of the cluster. In this scenario, the major part of the high
space velocity is simply the original velocity of the globular
cluster.

The ejection of the inner companion due to a dynamical instability in
a triple system, studied in detail by \citet{phln11} for
PSR\,J1903+0327, does not seem possible for PSR\,J1024$-$0719. The
reason is that the orbit of the outer companion (the K-dwarf/star B)
is expected to harden (i.e.\ become tighter) if the inner companion is
ejected. However, given the current extremely wide orbit, any release
of orbital binding energy would have been insufficient to
energetically eject the inner companion. The possibility that the
ejection event of the inner companion also caused the outer companion
to be perturbed into its current extremely wide orbit, from an
initially much closer orbit, is not allowed by energy considerations.

Detailed simulations of either the dynamically stable triple evolution
scenario presented here, or the globular cluster ejection scenario
suggested by \citet{kkn+16} will be required to ascertain which
scenario is most probable. When better constraints on the orbit of
PSR\,J1024$-$0719 become available in the future, predictions from
these scenarios can be tested.

\section{Discussion and conclusions}\label{sec:conclusions}
Since its discovery in 1994, PSR\,J1024$-$0719 was thought to be an
isolated millisecond pulsar. Our extended EPTA timing ephemeris of the
pulsar, combined with optical astrometry, photometry and spectroscopy
of the 2MASS\,J10243869$-$0719190 (star B), show that
PSR\,J1024$-$0719 is gravitationally bound to star B, and that the
gravitational interaction between the two can account for the apparent
discrepancy in the system distance and the steep spectrum timing noise
of PSR\,J1024$-$0719. We furthermore find that, even though we cannot
fully constrain the binary orbit, it must be extremely wide, with
orbital periods in excess of 200\,yrs. A rare formation scenario of
PSR\,J1024$-$0719 is required to explain the extremely wide binary
orbit and the very high space velocity in the Galaxy. We show that
PSR\,J1024$-$0719 could have evolved from a triple system, though
significant fine-tuning during the supernova explosion is required.

The binary orbit of PSR\,J1024$-$0719 and star B is presently not
fully constrained. As a result, the higher order frequency derivatives
in the radio timing solution of PSR\,J1024$-$0719 will remain free
parameters and will hence be a source of timing noise which may limit
using PSR\,J1024$-$0719 as a pulsar timing array source. To fully
constrain the orbit by timing PSR\,J1024$-$0719 alone, a measurement
of five spin frequency derivatives would be required
\citep{jr97}. Considering the component masses as known, and assuming
a value for either $\dot{P}_\mathrm{int}$ or $\ddot{z}_1$, the
frequency derivatives would constrain $P_\mathrm{b}$, $i$, $e$,
$\omega$ and $\nu$ and provide a description of the timing using the
spin frequency and spin frequency derivative as with other pulsars.

Improvements in the timing accuracy of PSR\,J1024$-$0719 are expected
in future IPTA data releases which combine EPTA, PPTA and NANOGrav
observations. The first IPTA data release \citep{vlh+16} contains
15.9\,yrs of EPTA and PPTA data on PSR\,J1024$-$0719. That release
contains a subset of the EPTA data used in this paper and hence does
not provide stronger constraints than those in
Table\,\ref{tab:ephemeris}. Combining the timing data presented here
with that of NANOGrav \citep{mnf+16,kkn+16} and the PPTA \citep{rhc+16}
should provide an increase in the timing accuracy which, to first
order, scales with the square root of the number of TOAs. The
uncertainties on the spin frequency derivatives with decrease over
time $t$ with $t^\alpha$, where $\alpha=-3/2$ for $f$, $-5/2$ for
$\dot{f}$, $-7/2$ for $\ddot{f}$ and so on. Hence, the uncertainties
on $f^{(3)}$ will already be halved by extending the timing baseline
another 4 years.

Independent astrometry of PSR\,J1024$-$0719 will be provided in the
near future through an ongoing VLBA pulsar astrometry project
\citep{dbc+11}. The expected uncertainties on position and proper
motions will be comparable to that from pulsar timing, while the
uncertainty on the pulsar parallax will be significantly better
(factor 5 to 10) compared to the current uncertainty of 0.11\,mas on
the timing parallax (Deller, priv.\,comm.).

An improvement in the position and proper motion of star B would also
help constrain the orbit through Eqns.\,\ref{eqn:pos} and
\ref{eqn:vel}. Though this would require including the longitude of
ascending node $\Omega$ as a free parameter, a measurement of
differential position and proper motion as well as two spin frequency
derivatives would allow a complete description of the orbit. Improving
the astrometry presented here through further ground-based
observations will be challenging. However, star B is bright enough for
its parallax, position and proper motion to be measured by
\textsc{Gaia}. The photometric transformations by \citet{jgc+10}
estimate a \textsc{Gaia} $G$-band magnitude of $G=19.1$ for its
$V$-band magnitude and $V-I$ color. The astrometric performance of
\textsc{Gaia}\footnote{\url{http://www.cosmos.esa.int/web/gaia/science-performance}}
predicts uncertainties on the parallax, position and proper motion of
0.32\,mas, 0.23\,mas and 0.17\,mas\,yr$^{-1}$, respectively. These
uncertainties are comparable to those of PSR\,J1024$-$0719 from the
timing solution.

Combining the \textsc{Gaia} astrometry of star B with the pulsar
timing astrometry of PSR\,J1024$-$0719 will provide much stronger
constraints on the orbit. For the nominal distance of $d=1.20$\,kpc,
these uncertainties translate to a distance uncertainty of 500\,pc, an
uncertainty in the projected position on the sky of 0.3\,AU and a
projected velocity uncertainty of 0.9\,km\,s$^{-1}$. Based on the
orbital solutions shown in Fig.\,\ref{fig:orbits}, the projected
velocities on the sky at the nominal distance are in the range of 0.3
to a few km\,s$^{-1}$, where the velocity is larger for less eccentric
orbits and with the binary members away from apastron, and hence may
be measurable from the astrometry of both binary members. This
approach will also provide independent constraints on the distance and
the intrinsic spin period derivative of PSR\,J1024$-$0719.

The full \textsc{Gaia} catalog is expected to be released in
2022. Given the prospect of much tighter constraints on the orbital
parameters with this future release, and considering the large pulsar
timing data set that has already been accumulated for
PSR\,J1024$-$0719, we consider it worthwhile to keep observing the
pulsar as part of pulsar timing arrays. Besides contributing to the
constraints on nano-hertz gravitational waves, these decade long,
multi-frequency, pulsar timing data sets like that presented here
enable us to measure other effects perturbing the stable rotation of
millisecond pulsars and make unexpected discoveries.

\section*{Acknowledgments}
We thank David Kaplan for fruitful discussions about the system and
the orbital constraints, which helped resolve a mistake in the
derivation of the spin frequency derivatives due to orbital motion.
This research has made use of the VizieR catalogue access tool, CDS,
Strasbourg, France. This research is based on observations made with
ESO Telescopes at the La Silla Paranal Observatory under programme ID
95.D-0973(A). Part of this work is based on observations obtained the
100-m telescope of the MPIfR (Max-Planck-Institut f\"ur
Radioastronomie) at Effelsberg. Pulsar research at Jodrell Bank and
access to the Lovell Telescope is supported by a Consolidated Grant
from the UK's Science and Technology Facilities Council. The
Nan\c{c}ay Radio Observatory is operated by the Paris Observatory,
associated with the French Centre National de la Recherche
Scientifique (CNRS). The Westerbork Synthesis Radio Telescope is
operated by the ASTRON (Netherlands Institute for Radio Astronomy)
with support from the Netherlands Foundation for Scientific Research
(NWO). The authors acknowledge the support of the collegues in the
European Pulsar Timing Array. CGB acknowledges support from the
European Research Council under the European Union's Seventh Framework
Programme (FP/2007-2013) / ERC Grant Agreement nr. 337062 (DRAGNET; PI
Jason Hessels). SO is supported by the Alexander von Humboldt
Foundation. KL acknowledges the financial support by the European
Research Council for the ERC Synergy Grant BlackHoleCam under contract
no. 610058.  RNC and PL acknowledge the support of the International
Max Planck Research School Bonn/Cologne and the Bonn-Cologne Graduate
School. GT and IC acknowledge financial support from `Programme
National de Cosmologie and Galaxies' (PNCG) of CNRS/INSU, France.

\bibliographystyle{mnras}

\bsp
\label{lastpage}

\onecolumn

\appendix
\section{Spin frequency derivatives due to orbital motion}\label{sec:orbit}
We start by expressing the orbit of a binary system on the sky in
terms of a projected separation $\rho$ and a position angle $\theta$:
\begin{equation}\label{eqn:pos}
\left( \begin{array}{c} x \\ y \\ z \end{array} \right) =
\left( \begin{array}{c} \rho \cos(\theta-\Omega) \\ \rho
  \sin(\theta-\Omega) \\ z \end{array} \right) =
\left( \begin{array}{c} r \cos(\omega+\nu) \\ r \sin(\omega+\nu) \cos
  i \\ r \sin(\omega+\nu) \sin i \end{array} \right).
\end{equation}
Here, $\nu$ is the true anomaly, $\omega$ the argument of perigee,
$\Omega$ the longitude of the ascending node and $i$ the
inclination. $z$ is the location along the line-of-sight, while $x$
and $y$ represent the position on the sky, but without correcting for
$\Omega$. We use the pulsar timing definition by \citet{dt92} and
\citet{kop96} which has positive $z$ away from the observer. As a
result, a right-handed system would have $\theta$ increase in a
clockwise direction. This is opposite to the more standard definition
of measuring position angles anti-clockwise on the sky, i.e.\ from
North to East. 

We follow \citet{fkl01} to obtain the velocities in the $x$, $y$ and
$z$ axes, which are defined as:
\begin{equation}\label{eqn:vel}
  \dot{x}=-\frac{2 \pi}{P_\mathrm{b}} \frac{a}{\sqrt{1-e^2}}(\sin(\omega+\nu)+e\sin\omega), 
\dot{y}=\frac{2 \pi}{P_\mathrm{b}} \frac{a \cos i}{\sqrt{1-e^2}}(\cos(\omega+\nu)+e\cos\omega),
\dot{z}=\frac{2 \pi}{P_\mathrm{b}} \frac{a \sin i}{\sqrt{1-e^2}}(\cos(\omega+\nu)+e\cos\omega).
\end{equation}
Here, $a$ is the semi-major axis and we have used that
\begin{equation}
  r=\frac{a(1-e^2)}{1+e\cos\nu},
  \dot{r}=\sqrt{\frac{G(m_1+m_2)}{a(1-e^2)}} e\sin\nu,
  r\dot\nu=\sqrt{\frac{G(m_1+m_2)}{a(1-e^2)}} (1+e\cos\nu),
  \left(\frac{2\pi}{P_\mathrm{b}}\right)^2=\frac{G(m_1+m_2)}{a^3}.
\end{equation}
Differentiating further for the radial position gives the
acceleration $\ddot{z}$, jerk $z^{(3)}$, snap $z^{(4)}$ and crackle
$z^{(5)}$:
\begin{eqnarray}
\ddot{z}=\left(\frac{2\pi}{P_\mathrm{b}}\right)^2 a \sin i\ q_2(e,\omega,\nu),
z^{(3)}=\left(\frac{2\pi}{P_\mathrm{b}}\right)^3 a \sin i\ q_3(e,\omega,\nu),
z^{(4)}=\left(\frac{2\pi}{P_\mathrm{b}}\right)^4 a \sin i\ q_4(e,\omega,\nu), z^{(5)}=\left(\frac{2\pi}{P_\mathrm{b}}\right)^5 a \sin i\ q_5(e,\omega,\nu).
\end{eqnarray}
where the $q_i(e,\omega,\nu)$ functions contain the contributions from
$e$, $\omega$ and $\nu$ and are:
\begin{eqnarray}
  q_2(e,\omega,\nu)=-\frac{(1+e\cos \nu)^2}{(1-e^2)^2}\sin(\omega+\nu),
  q_3(e,\omega,\nu)=-\frac{(1+e\cos \nu)^3}{(1-e^2)^{7/2}}[(1+e\cos\nu)\cos(\omega+\nu)-2e\sin\nu \sin(\omega+\nu)], \mathrm{and}\nonumber\\
  q_4(e,\omega,\nu)=\frac{1}{4}\frac{(1+e\cos \nu)^4}{(1-e^2)^5}[-4e\sin\omega+3e^2\sin(\omega-\nu)+4\sin(\omega+\nu)-6e^2\sin(\omega+\nu)+20e\sin(\omega+2\nu)+15e^2\sin(\omega+3\nu)],\\
  q_5(e,\omega,\nu)=\frac{1}{8}\frac{(1+e\cos \nu)^5}{(1-e^2)^{13/2}}[3e(9e^2-4)\cos\omega-15e^3\cos(\omega-2\nu)+10e^2\cos(\omega-\nu)+8\cos(\omega+\nu)-68e^2\cos(\omega+\nu)\nonumber\\
      +100e\cos(\omega+2\nu)-45e^3\cos(\omega+2\nu)+210e^2\cos(\omega+3\nu)+105e^3\cos(\omega+4\nu)]
\end{eqnarray}

These expressions give the line-of-sight time derivatives of position
of one binary member with respect to the other. In the case of pulsar
timing, these quantities are measured with respect to the binary
center of mass. Hence, the semi-major axis of the pulsar orbit $a_1$
is
\begin{equation}
a_1=a\frac{m_2}{m_1+m_2}
\end{equation}
where $m_1$ and $m_2$ are the pulsar and companion masses,
respectively. This allows us to define the orbital period as
\begin{equation}
\left(\frac{2\pi}{P_\mathrm{b}}\right)^2=\frac{k}{a_1^3}\ \mathrm{with}\ k=\frac{Gm_2^3}{(m_1+m_2)^2},
\end{equation}
which simplifies the time derivatives of $z$ with respect to the binary center of mass to
\begin{equation}\label{eqn:derivs}
  \ddot{z}_1=\frac{k\sin i}{a_1^2} q_2(e,\nu,\omega),
  z^{(3)}_1=\frac{k^{3/2}\sin i}{a_1^{7/2}} q_3(e,\nu,\omega),
  z^{(4)}_1=\frac{k^2\sin i}{a_1^5} q_4(e,\nu,\omega),\ \mathrm{and}\ z^{(5)}_1=\frac{k^{5/2}\sin i}{a_1^{13/2}} q_5(e,\nu,\omega)
\end{equation}

The time derivatives of $z_1$ translate into time derivatives of the
spin frequency $f$ through subsequent derivations of $\dot{f}=-f
\ddot{z}_1/c$, where $c$ is the speed of light. The derivatives yields
$\ddot{f}=-f z_1^{(3)}/c+\dot{f}^2/f$ and so on. We use the
approximation by \citet{jr97} by neglecting the higher order terms
which tend to be a factor $\dot{z}/c$ smaller and neglectable. As
such, we will use $\dot{f}=-f \ddot{z}_1/c$, $\ddot{f}=-f
z_1^{(3)}/c$, $f^{(3)}=-f z_1^{(4)}/c$ and $f^{(4)}=-f z_1^{(5)}/c$.

\end{document}